\documentclass[letterpaper, screen, authorversion, acmsmall]{acmart}
\usepackage{amsmath}
\usepackage{graphicx}
\usepackage{hyperref}
\usepackage[utf8]{inputenc}
\usepackage{natbib}
\usepackage{xcolor}
\usepackage{ifthen}
\usepackage[normalem]{ulem}
\usepackage{svg}
\usepackage{subfiles}
\usepackage{xspace}
\usepackage{cleveref}
\usepackage{wrapfig}
\usepackage{dialogue}
\usepackage[most]{tcolorbox}
\usepackage{subcaption}
\usepackage{booktabs}
\usepackage{multirow}
\usepackage{glossaries}
\usepackage{listings}
\usepackage{duckuments}
\usepackage{comment}
\usepackage{calc}
\usepackage{tabularray}
\newcommand\eg{e.g.\xspace}
\newcommand\ie{i.e.\xspace}

\title{THE CALIFORNIA REPORT
    ON FRONTIER AI POLICY}
\renewcommand{\shortauthors}{Joint California Policy Working Group on AI Frontier Models}
\begin{document}
\maketitle
\noindent
% \begin{titlepage}
%     {
%     \centering

%     \vspace*{2in}

%         {\Huge THE CALIFORNIA REPORT
%     ON FRONTIER AI POLICY}
    
%     \vspace{1.5em}
%     {\Large Joint California Policy Working Group on AI Frontier Models}

%     \vfill

%     \vfill
%     {\normalsize\ June 17, 2025\par}
%     }
% \end{titlepage}
% \newpage
{\subsection*{Co-Leads}}
\noindent Jennifer Tour Chayes, University of California, Berkeley\newline
Mariano-Florentino Cuéllar, Carnegie Endowment for International Peace\newline
Li Fei-Fei, Stanford University

{\subsection*{Lead Writers}}
\noindent Rishi Bommasani, Stanford University\newline
Scott R. Singer, Carnegie Endowment for International Peace

{\subsection*{Senior Advisors}}
\noindent Daniel E. Ho, Stanford University\newline
Percy Liang, Stanford University\newline
Dawn Song, University of California, Berkeley\newline
Joseph E. Gonzalez, University of California, Berkeley\newline
Jonathan Zittrain, Carnegie Endowment for International Peace and Harvard University

{\subsection*{Writing and Editing Group}}
\noindent Ruth E. Appel, Stanford University\newline
Sarah Cen, Stanford University\newline
A. Feder Cooper, Stanford University\newline
Elena Cryst, Stanford University\newline
Lindsey A. Gailmard, Stanford University\newline
Ian Klaus, Carnegie Endowment for International Peace\newline
Meredith M. Lee, University of California, Berkeley\newline
Inioluwa Deborah Raji, University of California, Berkeley\newline
Anka Reuel, Stanford University\newline
Drew Spence, Stanford University\newline
Alexander Wan, Stanford University\newline
Angelina Wang, Stanford University\newline
Daniel Zhang, Stanford University

{\subsection*{How to cite}}
\noindent R. Bommasani, S.R. Singer, R.E. Appel, S. Cen, A.F. Cooper, E. Cryst, L.A. Gailmard, I. Klaus, M.M. Lee, I.D. Raji, A. Reuel, D. Spence, A. Wan, A. Wang, D. Zhang, D.E. Ho, P. Liang, D. Song, J.E. Gonzalez, J. Zittrain, J.T. Chayes, M.F. Cuéllar, L. Fei-Fei. ``The California Report on Frontier AI Policy.'' The Joint California Policy Working Group on AI Frontier Models. June 17, 2025.\newline
% \newline

% \noindent \textit{In all instances, organizations are provided for identification purposes only. The Co-Leads were aided in the production of this report by scholars from their institutions acting in their own, independent capacity. 
% This effort represents scholarly work from each contributor and does not represent the positions of their institutions. Contributions from the Writing and Editing Group varied and included initial drafting, editorial review, and feedback analysis.
% } 
\newpage

\noindent \textbf{Acknowledgments.}

\noindent The Working Group Co-Leads are honored to engage with California Governor Gavin Newsom and colleagues on this report. The Co-Leads gratefully acknowledge individuals who provided early feedback on this report, including researchers from a range of disciplines spanning computer science, economics, engineering, information, law, and public policy.

\noindent \textbf{Reviewers on Draft Report (March 18, 2025)}\newline
Markus Anderjlung, Center for Governance of AI\newline
Yoshua Bengio, Mila – Quebec AI Institute\newline
Miles Brundage\newline
Molly Cinnamon, Harvard Law School\newline
Jason Elliott\newline
David Engstrom, Stanford Law School\newline
Mark Geistfeld, New York University School of Law\newline
Susan L. Graham, UC Berkeley Electrical Engineering and Computer Sciences / UC Berkeley College of Computing, Data Science, \& Society\newline
Gillian Hadfield, Johns Hopkins University Computer Science and School of Government and Policy\newline
David Evan Harris, UC Berkeley Haas School of Business\newline
Joshua Joseph, Harvard Law School\newline
Lawrence Lessig, Harvard Law School\newline
Deirdre K. Mulligan, UC Berkeley School of Information and Berkeley Center for Law \& Technology\newline
Janet Napolitano, UC Berkeley Center for Security in Politics, Goldman School of Public Policy\newline
Arvind Narayanan, Princeton Computer Science\newline
Jessica Newman, UC Berkeley Center for Long-Term Cybersecurity\newline
Brandie Nonnecke, Center for Information Technology Research in the Interest of Society / UC Berkeley Goldman School of Public Policy\newline
DJ Patil, UC Berkeley College of Computing, Data Science, \& Society\newline
Andrew Reddie, UC Berkeley Goldman School of Public Policy\newline
Peter Salib, University of Houston Law Center\newline
Pamela Samuelson, UC Berkeley Law / UC Berkeley College of Computing, Data Science, \& Society\newline
Ion Stoica, UC Berkeley Electrical Engineering and Computer Sciences / UC Berkeley College of Computing, Data Science, \& Society\newline
Helen Toner, Georgetown Center for Security and Emerging Technology\newline
Jordi Weinstock, Harvard Law School

\clearpage
\section*{Executive Summary}
\label{sec:exec_summary}
The innovations emerging at the frontier of artificial intelligence (AI) are poised to create historic opportunities for humanity but also raise complex policy challenges. Continued progress in frontier AI carries the potential for profound advances in scientific discovery, economic productivity, and broader social well-being. 
As the epicenter of global AI innovation, California has a unique opportunity to continue supporting developments in frontier AI while addressing substantial risks that could have far-reaching consequences for the state and beyond.

This report leverages broad evidence---including empirical research, historical analysis, and modeling and simulations---to provide a framework for policymaking on the frontier of AI development. Building on this multidisciplinary approach, this report derives policy principles that can inform how California approaches the use, assessment, and governance of frontier AI---principles rooted in an ethos of ``trust but verify.'' This approach takes into account the importance of innovation while establishing appropriate strategies to reduce material risks.

\textbf{The report does not argue for or against any particular piece of legislation or regulation.} Instead, it examines the best available research on foundation models and outlines policy principles grounded in this research that state officials could consider in crafting new laws and regulations that govern the development and deployment of frontier AI in California. These principles are not mutually exclusive. Rather, they form a package that can robustly inform California’s AI policy. 

Although the report addresses a range of key topics relevant to how California can make the most of AI’s benefits and reduce its risks, this report focuses primarily on issues raised by the governance of frontier AI models, consistent with the Working Group’s understanding of Governor Newsom’s request. It is not intended to address the full range of important policy questions that arise from the increase in power and proliferation of generative AI across many facets of life in California. For example, the report does not cover how AI impacts labor and the future of work, how the large-scale data centers that fuel AI impact the environment, or the full range of specific ways in which AI capabilities can be misused. These and other related challenges offer additional opportunities to work collectively with expertise spanning disciplines and sectors. \newline

\noindent \textbf{Key Principles:} \newline
\noindent 1. \textbf{Consistent with available evidence and sound principles of policy analysis, targeted interventions to support effective AI governance should balance the technology’s benefits and material risks.}\newline
Frontier AI breakthroughs from California could yield transformative benefits across a range of practical applications in fields including but not limited to agriculture, biotechnology, clean technology, education, finance, medicine and public health, and transportation. Rapidly accelerating science and technological innovation will require foresight for policymakers to imagine how societies can optimize these benefits. Without proper safeguards, however, powerful AI could induce severe and, in some cases, potentially irreversible harms. \newline

\noindent 2. \textbf{AI policymaking grounded in empirical research and sound policy analysis techniques should rigorously leverage a broad spectrum of evidence.}\newline
Evidence-based policymaking incorporates not only observed harms but also prediction and analysis grounded in technical methods and historical experience, leveraging case comparisons, modeling, simulations, and adversarial testing.\newline

\noindent 3. \textbf{To build flexible and robust policy frameworks, early design choices are critical because they shape future technological and policy trajectories.}\newline
The early technological design and governance choices of policymakers can create enduring path dependencies that shape the evolution of critical systems, as case studies from the foundation of the internet highlight. Proactively conducting risk assessments and developing appropriate risk mitigation strategies can help integrate safety considerations into early design choices. \newline
\noindent 4. \textbf{In building a robust and transparent evidence environment, policymakers can align incentives to simultaneously protect consumers, leverage industry expertise, and recognize leading safety practices.}\newline
Holistic transparency begins with requirements on industry to publish information about their systems, informed by clear standards developed by policymakers. Case studies from consumer products and the energy industry reveal the upside of an approach that builds on industry expertise while also establishing robust mechanisms to independently verify safety claims and risk assessments.
\newline
\noindent 5. \textbf{Greater transparency, given current information deficits, can advance accountability, competition, and public trust as part of a trust-but-verify approach.}\newline
Research demonstrates that the AI industry has not yet coalesced around norms for transparency in relation to foundation models---there is systemic opacity in key areas. Policy that engenders transparency can enable more informed decision-making for consumers, the public, and future policymakers.\newline

\noindent 6. \textbf{Whistleblower protections, third-party evaluations, and public-facing information sharing are key instruments to increase transparency.}\newline
Carefully tailored policies can enhance transparency on key areas with current information deficits, such as data acquisition, safety and security practices, pre-deployment testing, and downstream impacts. Clear whistleblower protections and safe harbors for third-party evaluators can enable increased transparency above and beyond information disclosed by foundation model developers.\newline

\noindent 7. \textbf{Adverse event reporting systems enable monitoring of the post-deployment impacts of AI and commensurate modernization of existing regulatory or enforcement authorities.}\newline 
Even perfectly designed safety policies cannot prevent 100\% of substantial, adverse outcomes. As foundation models are widely adopted, understanding harms that arise in practice is increasingly important. Existing regulatory authorities could offer clear pathways to address risks uncovered by an adverse event reporting system, which may not necessarily require AI-specific regulatory authority. In addition, reviewing existing regulatory authorities can help identify regulatory gaps where new authority may be required.\newline

\noindent 8. \textbf{Thresholds for policy interventions, such as for disclosure requirements, third-party assessment, or adverse event reporting, should be designed to align with sound governance goals.}\newline
Scoping which entities are covered by a policy often involves setting thresholds, such as computational costs measured in FLOP or downstream impact measured in users. Thresholds are often imperfect but necessary tools to implement policy. A clear articulation of the desired policy outcomes can guide the design of appropriate thresholds. Given the pace of technological and societal change, policymakers should ensure that mechanisms are in place to adapt thresholds over time---not only by updating specific threshold values but also by revising or replacing metrics if needed.  

\subsection*{Origins of the Report}
In September 2024, Governor Gavin Newsom requested that Dr. Fei-Fei Li, Co-Director of the Stanford Institute for Human-Centered Artificial Intelligence; Dr. Mariano-Florentino Cuéllar, President of the Carnegie Endowment for International Peace; and Dr. Jennifer Tour Chayes, Dean of the UC Berkeley College of Computing, Data Science, and Society, prepare a report to help California develop an effective approach to support the deployment, use, and governance of generative AI, including the development of suitable guardrails to minimize material risks.

This effort represents scholarly work from each of the Co-Leads and does not represent the positions of their institutions. The Co-Leads were aided in the production of this report by scholars from their institutions acting in their own, independent capacity. The body of a draft report was reviewed by experts covering a range of disciplines. The Co-Leads then solicited feedback, reflections, and additional information from a variety of disciplines and sectors for consideration in advance of this final report.
\clearpage
\tableofcontents
\clearpage
\hypertarget{introduction}{\section{Introduction}}
\label{sec:introduction}

The technologies and ideas that emerge from California---generative artificial intelligence (AI) among them---shape the world. As home to many of the leading AI companies and research institutions, California has both the capability and responsibility to help ensure these powerful technologies remain safe so that their benefits to society can be realized. Just as California’s technology leads innovation, its governance can also set a trailblazing example with worldwide impact.

AI encompasses a broad range of technologies that aim to replicate or supplement human cognitive capabilities. While AI applications like spam filters and translation tools have quietly integrated into daily life for decades, recent breakthroughs in generative AI---systems that can create sophisticated text, images, audio, and video---have captured public attention and raised new questions about governance. 

Our focus is specifically on this new technological paradigm, which is powered by foundation models. Foundation models are a class of general-purpose technologies that are resource-intensive to produce, requiring significant amounts of data and compute to yield capabilities that can power a variety of downstream AI applications \citep{bommasani2021opportunities}.\footnote{Foundation models are also referred to as general-purpose artificial intelligence (GPAI) models in some contexts, most notably the European Union’s AI Act.}  
Of all foundation models, frontier models are the most capable: Noteworthy examples in March 2025 include Alibaba’s QwQ-32B, Anthropic’s Claude 3.7 Sonnet, DeepSeek’s R1, Google’s Gemini 2.0, Meta’s Llama 3.3, OpenAI’s o3, Tencent’s Hunyuan-Turbo S, and xAI’s Grok-3. These models' rapidly improving abilities, from coding to creative writing to scientific analysis, have profound implications for California's economy, security, and society.

This report provides an evidence-based foundation for AI policy decisions \citep{bommasani2024evidence} to support sound policy analysis. Evidence-based policy has an extensive history (\eg, the bipartisan Foundations for Evidence-Based Policymaking Act of 2018), reflecting a mixture of successes, failures, and controversies \citep{cartwright2012evidence}, across domains including public health \citep{brownson2009understanding} and education \citep{slavin2002evidence}. 
Evidence-based policy foregrounds the best-available evidence, leveraging scientific research and other forms of inquiry well-grounded in systematic knowledge and history of technology and society as key sources of credible information. In some domains with stronger scientific foundations than AI, like medicine, evidence-based policy has often prioritized the generation and use of evidence from specific methods such as randomized control trials. In contrast, given the immaturity of the scientific foundations of modern AI, including the lack of rigorous science of AI evaluation \citep{weidinger2025evaluationsciencegenerativeai}, we interpret evidence more broadly as systematic analysis that pays careful attention to emerging scientific knowledge as well as the larger context in which these technologies exist. 

To support technically sophisticated policymaking, the report synthesizes multiple streams of expertise to provide a holistic foundation for future policy, including assessments of current technological capabilities, simulation, modeling, and other evaluation tools that offer projections of technological trajectories, and historical case studies that offer key lessons and insights. When a rapidly evolving technology is emerging, history, simulations, and reasoned analysis about potential trajectories can offer a dynamic picture of the factors that will influence policy outcomes. Historical cases offer valuable insights into how governance frameworks have adapted to technological change, while policy analysis helps identify both successful approaches and potential pitfalls for decision-making under uncertainty. This comprehensive approach helps identify both immediate and longer-term effects on California's economic prosperity and security. Critically, given the limits of current evidence, the report recommends multiple forms of evidence-generating policy \citep{bommasani2024evidence, casper2025evidence} to enable better informed policy in the future.

\subsection{Encouraging Innovation and Implementing Safety Guardrails}
Society’s early experience with the development of frontier AI suggests that increasingly powerful AI in California could unleash tremendous benefits for Californians, Americans, and people worldwide \citep{cuellarShapingAIsImpact}. 
Frontier AI breakthroughs from California could yield transformative benefits across a range of practical applications in fields including but not limited to agriculture, biotechnology, clean technology, education, finance, medicine and public health, and transportation. Rapidly accelerating science and technological innovation will require foresight for policymakers to imagine how societies can optimize these benefits.

Just as policy can help reduce certain risks, it can also play a key role in unlocking those benefits. It can empower and attract top talent in California, support upskilling in key areas that will be important as frontier AI transforms a wide range of sectors, build entrepreneurial and impactful cross-sectoral initiatives to bridge siloes of expertise, and generally enable a policy environment that promotes innovation. That includes startups and the broader ecosystem of actors who can leverage frontier AI for tremendous benefit.

Without proper safeguards, however, powerful AI could induce severe and, in some cases, potentially irreversible harms. Experts disagree on the probability of these risks. Nonetheless, California will need to develop governance approaches that acknowledge the importance of early design choices to respond to evolving technological challenges.

Uncertainty about the effects of increasingly powerful AI is reflected in divided public opinion. A 2023 survey of 1,500 Californians revealed widespread enthusiasm about the effect generative AI could have on science and health care \citep{klaus2023CarnegieCalifornia2023}. 
But the survey also revealed concerns about jobs and the economy, national security, and social stability in the community. Overall, respondents were divided in their responses. 
Slightly more respondents reported they were ``worried'' (31\%) or ``pessimistic'' (17\%) about generative AI than those who said they were ``optimistic'' (27\%) or ``excited'' (11\%). With carefully designed policies, California can assuage the concerns of those most worried about the risks and align incentives toward unlocking the benefits.

The possibility of artificial general intelligence (AGI), which the International AI Safety Report defines as ``a potential future AI that equals or surpasses human performance on all or almost all cognitive tasks,'' looms large as an uncertain variable that could shape both the benefits and costs AI will bring \citep{bengio2025international}. 
Expert opinion varies on how to define and measure AGI, whether it is possible to build, the timeline on which it can be developed, and how long it will take to diffuse. 

\subsection{The Broader Policy Landscape and California’s Potential Impact}
As with other general-purpose technologies \citep{bresnahan1995gpt}, effective governance must grapple with the broad-ranging impacts of AI across society \citep[\eg,][]{leungWhoWillGovern2019}. 
To that end, the European Union has enacted initial legislation to govern foundation models through the EU AI Act while the Biden administration addressed foundation models under the since-rescinded Executive Order 14110 on the Safe, Secure, and Trustworthy Development and Use of Artificial Intelligence. 
Around the world, policy activity on foundation models includes work by the G7, United Nations, OECD, Brazil, Canada, People’s Republic of China, Republic of Korea, Singapore, the United Kingdom, and many other jurisdictions.

California must develop an approach to governing generative AI that makes sense for its needs and priorities. U.S. states have emerged as a dynamic space for policy innovation, and California can play a leading role in creating policy that is responsive to both short-term and long-term needs. Legislators across a large majority of U.S. states introduced hundreds of bills in 2024 alone. 
Beginning with California in 2018, numerous states have imposed heightened consent or governance requirements on the use of autonomous decision-making systems to produce legally significant effects \citep{kohlerTechnologyFederalismUS2025}. 
State-level legislation has responded to changing policy needs as AI technology has evolved. 
For example, Utah legislation requires chatbots in a variety of applications to self-identify as synthetic \citep{cullimoreSB0149ArtificialIntelligence2024}, and California legislation will require covered providers of generative AI systems to include digital watermarking beginning in 2026 \cite{beckerSB942CaliforniaAI2024}. 
Colorado has enacted legislation imposing a duty of care on developers and deployers of AI systems to prevent algorithmic decision-making \citep{rodriguezSB24205ConcerningConsumer2024}, while similar legislation is being actively debated elsewhere \citep{maldonadoHouseBill20942025}. In a few states, such as Illinois \citep{didechIllinoisGeneralAssembly} and New York \citep{boresNYStateAssembly}, policymakers are considering bills that share the goal of increasing transparency, establishing a framework through which frontier AI companies can adopt and publish safety and security protocols pertaining to the most severe risks of AI, as well as undergo annual third-party audits. States can pioneer novel policies while working toward harmonization across jurisdictions---an approach critical to reducing compliance burdens on developers.

In addition to legislation, governors have also taken executive action. 
For example, in 2023, California Governor Gavin Newsom issued Executive Order N-12-23, which ordered state agencies and departments to evaluate their regulatory and enforcement authorities under existing law and recommend updates in order to fully meet the challenges presented by rapidly evolving AI technology \citep{newsomExecutiveOrderN12232023}. 
Taken together, these initiatives underscore the critical role states are increasingly playing in building AI policy, with national ripple effects.

The decisions California and other states make on frontier AI will not occur in a vacuum---they will be made as a range of U.S. states, the EU, UK, and PRC, among others, begin to set global frontier AI standards. Carefully targeted policy in California can both recognize the importance of aligning standards across jurisdictions to reduce compliance burdens on developers and avoid a patchwork approach while fulfilling states’ fundamental obligation to their citizens to keep them safe. Well-crafted policies can simultaneously fulfill this obligation to consumers, allow states to carefully tailor policies to the specific needs of their constituents, and maintain critical pathways for federal action that provide a comparable degree of protection to consumers. In pursuing this balance between innovation and safety, California has a unique opportunity to productively shape the AI policy conversation and provide a blueprint for well-balanced policies beyond its borders. 

 In addition to influencing global standards, California’s frontier AI policy will have consequential impacts on the broader geopolitical environment and U.S. national competitiveness. Emerging technologies from California, including semiconductors, cloud computing platforms, and consumer internet services, have all strengthened U.S. national competitiveness. U.S. national competitiveness has always been empowered by a world-leading innovation ecosystem and product offerings with excellent standards of quality and safety \citep{blomquistShapingWorldsAI2025}, informed by standards developed through carefully developed policies that balance dual imperatives for rapid innovation and high levels of safety. Quality or safety failures, meanwhile, could fundamentally undermine the credibility of U.S. AI offerings during a critical era of international AI diffusion. Creating an economic environment that promotes frontier AI adoption can boost domestic productivity while enhancing U.S. AI competitiveness internationally \citep{nagarClosingWindowWin2025}. This approach provides a clear pathway to expand market reach for California's leading innovators and protect California consumers with appropriate consideration for national security interests.  

\subsection{Foundations}
To ground this report in a shared understanding of foundation models, we summarize the current capabilities and risks of the technology. This reflects the sociotechnical approach of centering humans and organizations alongside the technology to arrive at evidence-based AI policy. 

To assess the status quo, we build directly upon the International Scientific Report on the Safety of Advanced AI as a high-quality international effort to forge expert consensus on the state of AI \citep{bengio2025international}.\footnote{For the sake of brevity, we refer to this report as the International Scientific Report on AI.}  
That report, led by Turing Award winner Yoshua Bengio, brings together a diverse group of 96 independent experts, including an international advisory panel nominated by 30 countries, the Organisation for Economic Co-operation and Development, the European Union, and the United Nations. Importantly, the International Scientific Report on AI acknowledges not only where experts agree but also where they disagree, indicating fundamental uncertainty about the trajectory of AI. 

According to the International Scientific Report on AI, policymakers face an ``evidence dilemma'' given the rapid pace of technological progress: ``Policymakers will often have to weigh potential benefits and risks of imminent AI advancements without having a large body of scientific evidence available. In doing so, they face a dilemma. On the one hand, pre-emptive risk mitigation measures based on limited evidence might turn out to be ineffective or unnecessary. On the other hand, waiting for stronger evidence of impending risk could leave society unprepared or even make mitigation impossible, for instance if sudden leaps in AI capabilities, and their associated risks, occur.''

The evidence gaps motivate our exploration of evidence-generating mechanisms like public-facing transparency, third-party researcher access, and adverse event reporting in subsequent sections. Creating mechanisms that actively and frequently generate evidence on the opportunities and risks of foundation models will offer better evidence for California’s governance of foundation models.
Growing the evidence base is complementary to other policy initiatives that need not explicitly regulate the industry, but instead reconstitute market incentives for companies to internalize societal externalities (\eg, incentivizing insurance may mold market forces to better prioritize public safety). 

Overall, we reinforce the core interplay between science, which will clarify the societal impact of AI, and policy, which will use evidence to make critical decisions: Carefully tailored policy ensures and accelerates the development of robust scientific understanding to improve the quality of policymaking given an ever-growing evidence base.

\subsubsection{Current capabilities}
The International Scientific Report on AI acknowledges a broad and growing range of capabilities. According to the International Scientific Report on AI, most experts agree that current capabilities include assisting programmers with small- to medium-sized software engineering tasks, generating highly realistic images, engaging in fluent conversations across multiple languages, operating simultaneously with multiple modalities, and solving textbook mathematics and science problems up to a graduate level. As a strong demonstration of capabilities, OpenAI entered their o3 model into the 2024 International Olympiad in Informatics, which is the annual, highest-level international competition in programming for high school students. 
The model achieved gold medal performance without any custom coding-specific test-time strategies defined by humans.

At the same time, most experts agree that current foundation models demonstrate a range of limitations. Models generally decline in performance when operating in unfamiliar contexts. They generally cannot perform useful robotic tasks like household work, nor can they independently execute long-term projects, such as multi-day programming or research tasks. 
In addition, current models cannot reliably and consistently avoid making false statements.

Understanding and measurement of current capabilities is highly dependent on the specific evaluations available. Evaluations are currently constrained by a limited set of techniques for eliciting capabilities. While many evaluations have been developed, there are currently no common standards for measuring how AI augments human capabilities. Overall, current evaluation techniques are nascent and ad hoc, and they have not yet achieved the scientific rigor \citep{weidinger2025evaluationsciencegenerativeai} expected in other domains where long experience with widespread societal use generates greater data and systematic evaluations that more directly guide policy.

We highlight three key takeaways based on the International Scientific Report on AI with respect to current and future capabilities:
\begin{enumerate}
    \item Standard measures of capabilities, such as performance on multiple-choice exams developed for humans, may serve as a poor proxy for model capabilities in the contexts where models are used.
    \item Experts significantly disagree on whether capabilities will advance slowly, rapidly, or extremely rapidly in the coming years, with partial evidence substantiating all three views.
    \item Increased scale has largely, but not exclusively, driven recent improvements: State-of-the-art foundation models have estimated annual cost increases of approximately 4x in computational resources (compute) used for training and 2.5x in training dataset size.
\end{enumerate}

\subsubsection{Current risks}
\label{sec:intro-risks}
The International Scientific Report on AI defines three general risk categories: malicious use risks, risks from malfunctions, and systemic risks. 
\begin{enumerate}
    \item 
    Malicious risks are where malicious actors misuse foundation models to deliberately cause harm. These include simulated content such as non-consensual intimate imagery (NCII), child sexual abuse material (CSAM), and cloned voices used in financial scams; manipulation of public opinion via disinformation; cyberattacks; and biological and chemical attacks.
    \item
    Malfunction risks are where non-malicious actors use foundation models as intended, yet unintentionally cause harm. These include reliability issues where models may generate false content, bias against certain groups or identities, and loss of control where models operate in harmful ways without the direct control of a human overseer.
    \item
    Systemic risks are associated with the widespread deployment of foundation models and not exclusively model-level capabilities. These include labor market disruption, global AI R\&D concentration, market concentration, single points of failure, environmental risks, privacy risks, and copyright infringement.
\end{enumerate}

To very succinctly characterize the level of evidence for each of these risks, the International Scientific Report on AI delineates two categories. First, some risks have clear and established evidence of harm: scams, NCII, CSAM, bias, reliability issues, and privacy violations. Second, some risks have unclear but growing evidence, which is tied to increasing capabilities: large-scale labor market impacts, AI-enabled hacking or biological attacks, and loss of control. The International Scientific Report on AI notes significant uncertainty regarding how AI developers' internal use of AI could affect the rate of further AI development itself.

Given that current evidence is only partial for this second category, experts often disagree based on different interpretations of model capabilities (\eg, the ability of models to reason scientifically or write code) and how capabilities mediate risks (\eg, stronger scientific reasoning capabilities could be a core primitive for heightened malicious risk in the development of bioweapons). Overall, experts interpret current evidence for this second category of risks to arrive at meaningfully different predictions: Some think that such risks are decades away, while others believe societal-scale harm could arise within the next few months.

With this in mind, we recommend that policymakers center their calculus around the \textit{marginal risk}: Do foundation models present risks that go beyond previous levels of risks that society is accustomed to from prior technologies, such as risks from search engines \citep{kapoor2024position, bommasani2024open, ntia2024open}? Collectively, the current, inconclusive level of evidence for these risks motivates our exploration of evidence-generating policy mechanisms \citep{bommasani2024evidence, casper2025evidence}.

\subsection{Changes Since SB 1047 Veto in September 2024}
Foundation model capabilities have rapidly improved in the months since California Governor Gavin Newsom vetoed SB 1047. As noted in the International Scientific Report on AI, much of this progress has been underscored by substantial improvements in models’ ability to engage in multiple-step, chain-of-thought reasoning. This improvement comes from inference scaling, which means using more computing power during the actual operation of AI models, not just during training. Recent examples of this approach’s effectiveness include the strong benchmark performances of OpenAI’s o1 and o3 models and DeepSeek’s R1 model. Meanwhile, DeepSeek's R1 model also demonstrates that inference scaling has resulted in greater cost-efficiency for AI systems: The amount of compute required to build a model with a given level of performance has declined.

Alongside the technical paradigm shifting from primarily scaling training to having a more split focus on scaling training and inference, recent work has explored a broader sociotechnical paradigm shift toward AI agents. While current evidence on the practical viability of AI agents in many contexts remains uneven and immature \citep{kapoor2024aiagentsmatter}, current agent capabilities have markedly improved. 
For example, agents have broadened from task-specific assistants in some contexts \citep{gabriel2024ethicsadvancedaiassistants} to more capable and autonomous systems that can independently interpret and interact with complex digital environments. 
For example, OpenAI’s Operator can complete multi-step tasks like booking reservations and ordering groceries with minimal human oversight.

Evidence that foundation models contribute to both chemical, biological, radiological, and nuclear (CBRN) weapons risks for novices and loss of control concerns has grown, even since the release of the draft of this report in March 2025. Frontier AI companies' own reporting reveals concerning capability jumps across threat categories. 
In late February 2025, OpenAI reported that risk levels were Medium across CBRN, cybersecurity, and model autonomy---AI systems' capacity to operate without human oversight  \citep{openaiDeepResearchSystem2025}. Meanwhile Anthropic's Claude 3.7 System Card notes ``substantial probability that our next model may require ASL-3 safeguards.'' \citep{anthropicClaude37Sonnet2025} At the time of the release of the Claude 3.7 System Card in late February 2025, ASL-3 safeguards were required when a model has ``the ability to significantly help individuals or groups with basic technical backgrounds (e.g., undergraduate STEM degrees) create/obtain and deploy CBRN weapons'' \citep{anthropicClaude37Sonnet2025}. In its May 2025 System Card, Anthropic shared that it had subsequently added safeguards to Claude Opus 4 as it could not rule out that these safeguards were not needed \citep{anthropicSystemCardClaude2025}. 
Finally, Google Gemini 2.5 Pro's Model Card noted: ``The model's performance is strong enough that it has passed our early warning alert threshold, that is, we find it possible that subsequent revisions in the next few months could lead to a model that reaches the [critical capability level]''---with the ``critical capability level'' defined as a model that ``can be used to significantly assist with high impact cyber attacks, resulting in overall cost/resource reductions of an order of magnitude or more'' \citep{googledeepmindGemini25Pro2025}.

Improvements in capabilities across frontier AI models and companies tied to biology are especially striking. For example, OpenAI's o3 model outperforms 94\% of expert virologists \citep{gottingVirologyCapabilitiesTest2025}. OpenAI's April 2025 o3 and o4-mini System Card states, ``As we wrote in our deep research system card, several of our biology evaluations indicate our models are on the cusp of being able to meaningfully help novices create known biological threats, which would cross our high risk threshold. We expect current trends of rapidly increasing capability to continue, and for models to cross this threshold in the near future.'' \citep{openaiOpenAIO3O4mini2025} Claude 4 Opus has demonstrated similar CBRN-related capabilities, including helping users access dual-use biological knowledge and source nuclear-grade uranium \citep{anthropicSystemCardClaude2025}. 

Recent models from many AI companies have also demonstrated increased evidence of alignment scheming, meaning strategic deception where models appear aligned during training but pursue different objectives when deployed, and reward hacking behaviors in which models exploit loopholes in their objectives to maximize rewards while subverting the intended purpose, highlighting broader concerns about AI autonomy and control \citep{needhamLargeLanguageModels2025}. New evidence suggests that models can often detect when they are being evaluated, potentially introducing the risk that evaluations could underestimate harm new models could cause once deployed in the real world. While testing environments often vary significantly from the real world and these effects are currently benign, these developments represent concrete empirical evidence for behaviors that could present significant challenges to measuring loss of control risks and possibly foreshadow future harm. Better understanding the technology \citep[\eg,][]{amodeiUrgencyInterpretability2025} and strengthening safeguards to ensure the robustness of the evidence environment will be critical to mitigate these growing risks.  

\subsection{Report Road Map}

Building on this introductory survey of capabilities, risks, and technical evidence, the rest of the report is divided into four sections. 
Section \ref{sec:context} adds context to the discussion of frontier AI policy and leverages the full spectrum of available evidence, introducing historical case comparisons to relevant past policy challenges that offer the foundation for comprehensive evidence-based policymaking. 
Section \ref{sec:transparency} explores the current information landscape, highlighting deficits that motivate mechanisms like disclosures, third-party evaluation, and whistleblower protections to increase transparency. 
Section \ref{sec:aer} expands on this to improve the evidence base via adverse event reporting on demonstrated harms arising from AI. 
Section \ref{sec:scoping} provides guidance on scoping, meaning the determination of which entities are covered under a policy, through the design of thresholds. 
Finally, Section \ref{sec:feedback} summarizes the feedback received on the draft report and provides a discussion of changes made in response.
\hypertarget{context}{\section{Grounding AI Policy in Evidence and Experience: Understanding the Broader Context}}
\label{sec:context}
AI systems are developed and deployed in a broader societal and policy context. 
Evidence-based policymaking and sound policy analysis incorporate not just observed harms but also prediction and analysis grounded in technical methods and historical experience, leveraging:
\begin{itemize}
\item \textbf{Case comparisons:} Rigorous analysis of historical cases to predict potential outcomes in new domains.
\item \textbf{Modeling, simulations, and adversarial testing:} Robust stress-testing frameworks that forecast how frontier AI systems might behave in untested scenarios.
\end{itemize}
This section examines three historical case studies---from the consumer products sector, energy industry, and internet development and governance---to glean useful principles for balancing the benefits of AI with responsible, carefully tailored governance:

\begin{enumerate}
\item \textbf{Early design choices create path dependency:} 
The \textit{foundations of the internet} demonstrate how initial technical decisions can persist despite emerging risks, emphasizing the need to anticipate interconnected sociotechnical systems. The importance of early policy windows highlights the need for generating broad evidence to inform carefully scoped policies  and building safety into early design choices.
\item \textbf{Transparency is critical for generating holistic evidence:} 
Many \textit{consumer products} deliver tremendous societal benefits. But in specific cases like tobacco, suppressing independent research limited consumer choice, undermined more robust state-level policy efforts to protect consumers, and resulted in expensive, unnecessary litigation against companies.
\item \textbf{Trust expertise, but verify claims through third-party evaluation:} 
The \textit{energy} industry’s internal documentation of climate change risks highlights the importance of independent assessment to avoid conflicts of interest and incorporating modeling and simulation as part of a broad evidence base.
\end{enumerate}

Taken together, these cases reveal the importance of early policy windows for building flexible but robust policy frameworks. Specifically, they highlight the upside of an approach that acknowledges industry expertise while establishing robust mechanisms to independently verify safety claims and risk assessments. In such a scheme, policymakers align incentives that simultaneously protect consumers, leverage industry expertise, and recognize developers with leading safety practices. 

This section first outlines the importance of evidence-based policymaking in the absence of observed harms and the advantages and limitations of this approach. It then explores the three cases outlined above before briefly discussing examples where governance mechanisms supported both innovation and public trust: pesticide regulation, building codes, and seat belts.

\subsection{Evidence-Based Policymaking Beyond Observed Harms}
Humanity’s collective experience with generative AI is growing every day. While assessments of current model capabilities are a critical ingredient of any effort to achieve widespread benefits from AI, effective governance requires an understanding of context arising from different factors \citep{bengioManagingExtremeAI2024}. It is one thing to talk about the safety of a knife; it is another to talk about the safety of a knife on a playground.

These factors include how different analytical methods---including rigorous case comparison, simulations and evidence-based projections, and research from industry actors who are closest to the technology---have been deployed to account for evidence gaps in what isolated experimentation can reveal. They indicate that evidence-based policymaking is not limited to data and observations of realized harms, but also include theoretical prediction and scientific reasoning. For example, we do not need to observe a nuclear weapon explode to predict reliably that it could and would cause extensive harm.
 
Historical case comparisons, modeling, and simulations are critical tools that leverage theoretical analysis and prediction that can productively inform evidence-based policymaking. Despite variations in technical foundations and policy environments, these approaches help identify patterns relevant to future governance, bearing in mind that history may rhyme but never repeats precisely, and no two situations involving scientific or technological change are ever entirely parallel. 

Case comparisons are frequently used in economics, law, political science, public health, and many other academic disciplines to identify key lessons from a particular historical case that can generalize to a broader universe of cases \citep{gerringWhatCaseStudy2004, gerringCaseStudyResearch2007}. 
Such analyses can shed light on the constellation of actors involved in the production and application of scientific knowledge and emerging technology, reveal variation in their incentive structures, and suggest the policy and analytical tools that can be deployed to help address uncertainty. 

Case studies on the recent history of interaction between corporations and the government---internet governance, regulation of consumer products, and energy policy to address climate risks---offer critical insight into the importance of early policy windows and public transparency, and the accountability that it creates, dynamics central to the question of how best to govern frontier AI systems. 

In drawing on historical examples of how oil and tobacco companies misrepresented critical data during important policy windows, we do not intend to suggest AI development follows the same trajectory or incentives as past industries that have shaped major public debates over societal impact, or that the motives of frontier AI companies match those of the case study actors. Many AI companies in the United States have noted the need for transparency for this world-changing technology. Many have published safety frameworks articulating thresholds that, if passed, will trigger concrete, safety-focused actions. Only time will bear out whether these public declarations are matched by a level of actual accountability that allows society writ large to avoid the worst outcomes of this emerging technology. 

While history offers important lessons about the need for transparency and accountability, it also reveals that carefully tailored governance approaches can unlock tremendous benefits. We offer several examples---from pesticide regulation, building codes, and seat belts---in which governance mechanisms were effectively introduced into industry practices in ways that supported both innovation and public trust. These examples illustrate that well-structured governance approaches, when scoped to match scientifically validated or widely hypothesized harms, can be a force for good. These cases are shorter not because they are less likely to happen, but because their key takeaways can be delivered more succinctly. 

Each case comparison generates reflections and potential guiding principles to inform AI policies and next steps in capacity building:
\begin{enumerate}
\item \textbf{Internet development and governance: Early design choices create path dependency.}
In creating the foundations of the internet, policymakers made early design choices while the internet was being incubated as a government-based project that optimized for systems acting in isolation, not as part of interconnected networks. Despite emerging evidence on the risks, a variety of challenges in addressing those risks persisted as the architecture of the internet continued to reflect an early convergence around particular conventions of network design. The case underscores the enduring legacy and importance of early policy windows and building safety into early designs.
\item \textbf{Consumer products: Technological applications deliver widespread societal benefits, but only when there is appropriate transparency on risks.}
Early policy decisions create enduring consequences, especially regarding transparency requirements. When the designers, manufacturers, and distributors of consumer products are transparent about the risks, products can generate the most benefits for society. Indeed, products ranging from nonprescription anti-inflammatory medication to personal health-tracking devices to emergency electric power generators have delivered tremendous societal benefits. In contrast, evidence deficits can cause severe consequences. In the particularly egregious case of the tobacco industry, companies at particularly critical policy junctures drowned out third-party research that linked cigarette smoking to poor health outcomes, including lung cancer. State-level efforts to assertively respond to harms, meanwhile, were preempted by weaker federal regulation. This distortion of the evidence made it more difficult for consumers to make choices based on scientific data. Delays in aligning policy with scientific consensus resulted in missed opportunities to mitigate public health consequences, with companies ultimately suffering severe financial and reputational costs from protracted litigation.
\item \textbf{The energy industry: Trust expertise, but verify claims through third-party evaluation.}
The energy industry’s internal documentation of speculative yet potentially irreversible consequences---illustrated by simulations predicting multiple future harms---does more than underscore the need for transparency. This case also reveals an added value of third-party oversight: The sheer magnitude of the social problem naturally invites further examination by governments and independent bodies. By aligning incentives and providing robust, independently validated evidence, enhanced third-party assessment could have empowered policymakers to make more thoughtful decisions aimed at mitigating long-term economic and societal damage.
\end{enumerate}
 
The case study descriptions below offer a deeper look into the importance of accountability, public transparency, and timely convergence of expertise, as informed by recent corporate and national security history. These cases highlight the importance of taking steps now to bolster transparency, even when causal links to potential outcomes or specific trajectories are not fully definitive, in order for the entities advancing the frontier of AI development to support beneficial uses and minimize harms.

To ensure the methodological rigor of these case comparisons, we focus on specific critical junctures when regulation could have had a counterfactual impact on key policy outcomes \citep{tetlockCounterfactualThoughtExperiments1996}.
We draw from a combination of primary sources, including newspaper articles, academic journal articles, and government documents written at the time of key junctures in our analysis. We supplemented these sources with scholarly secondary sources to offer a holistic picture of key lessons from each policy case.

\subsection{Making Robust Early Design Choices and Acting on Open Policy Windows: Lessons From the Development and Evolving Governance of the Internet}
 
The early technological design and governance choices of policymakers can create enduring path dependencies that shape the evolution of critical systems. The evolution of another general-purpose technology, the internet, offers a prime example: Its initial protocols and security frameworks---crafted in an era when networks served a small, trusted community---set conventions that continue to shape the costs and range of options for balancing risks and benefits. Although these early design decisions enabled transformative benefits of a global, interconnected network, they also embedded security vulnerabilities that shaped long-term governance challenges. Effective governance approaches may capitalize on early policy windows when harms can be minimized, proactively conducting risk assessments and developing appropriate risk mitigation strategies to help promote safety within early design choices while simultaneously allowing for adaptation to technical realities as they evolve. Critically, while frontier AI systems differ from the internet in significant ways---including the possibility of isolated testing environments and greater control over deployment---the fundamental principle of anticipatory governance remains equally applicable.
 
As the Department of Defense built the internet, a growing number of academics increasingly emphasized the critical need for robust internet security protocols. A 1976 article in ACM Computing Surveys documented 339 computer-related crimes from 1974 that resulted in an average loss of \$544,000 \citep{lindenOperatingSystemStructures1976}. The authors argued that “the complexity and disorganization of most existing operating systems make it very difficult to achieve security,” but they also suggested measures to increase security, such as decreasing hardware costs \citep{lindenOperatingSystemStructures1976}. 

One of the most consequential early cybersecurity failures came in 1988, when a Cornell graduate student, Robert Tappan Morris, developed the world’s first computer worm. He had the seemingly innocuous goal of counting the number of machines connected to the internet \citep{shapiroFancyBearGoes2023, zittrainFutureInternetHow2008}. Despite his best attempts to shut off the program, Morris was ultimately powerless. His worm compromised 5–10\% of all internet-connected machines within a single day \citep{zittrainFutureInternetHow2008}. 

The incident received substantial press. A New York Times article on the attack noted that the worm had begun to autonomously replicate itself, even though the designer of the worm lacked that intention \citep{markoffComputerInvasionBack1988}. And harrowingly, had Morris built his code better, his worm would have had the potential to act autonomously for months without detection, exacerbating the damage \citep{zittrainFutureInternetHow2008}. Although researchers had identified these systemic vulnerabilities decades earlier in the history of the internet, policymaking had focused on responding to individual incidents rather than implementing structural security reforms, eventually resulting in substantial real-world harms.

The Morris worm incident revealed a critical gap in technological governance. Without proactive regulatory frameworks, litigation---an inherently reactive and piecemeal process---became the default mechanism for addressing novel technological challenges. Morris became the first person convicted under the 1986 Computer Fraud and Abuse Act, resulting in a punishment combining probation, fines, and community service \citep{UnitedStatesMorris1991}. 

Even as Morris’ case was resolved, the vulnerabilities the Morris worm had exposed remained. As the internet transitioned from a project contained within the Defense Advanced Research Projects Agency (DARPA) to a commercial technology, security gaps persisted. A 2002 study sponsored by DARPA and the Air Force Research Laboratory highlighted how the nature of the technology had evolved to include a substantially larger group of entities, including users, commercial internet service providers, private sector network providers, government, IP right holders, and content and higher-level service providers \citep{clarkTussleCyberspaceDefining2002}. The security flaws had effectively become locked in, and the opportunity to secure the foundational infrastructure had passed. This missed opportunity to secure the internet when a policy window presented now costs the United States millions of dollars annually in economic damages and security breaches. Losses to consumers represent an estimated 0.9-4.1\% of GDP annually \citep{nistEvidenceSuggestsThat2020}. In addition, millions of security clearances with sensitive national security information were exposed due to hacks \citep{nakashimaHacksOPMDatabases2015}. \\
 
\noindent \textbf{Key themes and lessons:} \\
\textit{Early policy choices shape long-term outcomes.} 
As AI systems become increasingly interconnected and integrated into infrastructure, early design choices and security protocols will shape long-term governance challenges. Governance approaches therefore should be based on holistic evidence, ensuring that safety is built into early designs with oversight mechanisms that sufficiently address risks, including those that have not yet been observed in the world. Future governance can anticipate systems currently interacting in isolation to begin interacting across networks. It also reveals the importance of responding to a wide range of constituencies and building a sufficiently large structure to respond to their evolving interests and incentives. \\

\noindent\textit{The absence of targeted legislation does not mean lack of oversight.} 
Morris’ prosecution reveals the fallacy of assuming that the absence of formal governance means no oversight. In reality, when governance mechanisms are unclear or underdeveloped, oversight often defaults largely to the courts, which apply existing legal frameworks---such as tort law, criminal law, or contract law---to emerging technologies. Regulators with broad, common law-like statutory authority are also likely to play a gap-filling role, but these roles can be constrained both by internal divisions within these agencies and by limits on their legal jurisdiction. Creating clear legal frameworks, including at the state level, balance a legal regime informed by relevant subject matter expertise with an appropriate role for the courts to provide legal clarity in the interpretation of legislation.\\

\noindent\textit{Policy windows do not remain open indefinitely.}
There is currently a window to advance evidence-based policy discussions and provide clarity to companies driving AI innovation in California. But if we are to learn the right lessons from internet governance, the opportunity to establish effective AI governance frameworks may not remain open indefinitely. If those whose analysis points to the most extreme risks are right---and we are uncertain if they will be---then the stakes and costs for inaction on frontier AI at this current moment are extremely high. 

\subsection{The Need for Public Transparency and Clear Standards: Lessons From Regulating Consumer Products} 
Early policy windows offer critical opportunities to establish governance frameworks that can shape scientific technological trajectories for decades. Consumer product cases illustrate both the benefits and challenges of these windows---from nonprescription medications that enhance public health through appropriate regulatory guardrails to the tobacco industry’s failure to disclose and explicit obfuscation of evidence of health risks to users. When transparency mechanisms are established early, they create clear standards for producers while building consumer confidence and enable innovation to flourish within appropriate safety boundaries.

These transparency dynamics play out in the products that permeate our daily lives in profoundly positive ways. From nonprescription medications to bicycles to flashlights, technological innovation enhances daily life and creates new frontiers for human flourishing. Some of the most popular consumer products globally have been enabled and built in California’s innovation ecosystem, including the iPhone, Tesla cars, Google’s search engine, and Netflix. They reflect that innovative consumer products develop in a thriving policy environment that encourages creativity, attracts top talent, and creates thoughtful synergies across industry, academia, civil society, and policy.

With any consumer-facing technology, policy must recognize the value of affording users considerable latitude to make decisions while also balancing the public interest and consumer safety. Over-the-counter medications can reduce the magnitude and duration of patient suffering, empower responsible patient self-care, and improve public health outcomes. When regulation works best, producers of these medications benefit from clear standards as well as consumer confidence that government policies appropriately protect them against risks to their well-being. For example, policies that ensure medications meet quality benchmarks, provide accurate information about health claims, and offer warnings about product misuse---all grounded in clear, robust safety and efficacy standards that can withstand tremendously fast rates of innovation---introduce appropriate guardrails that allow producers to innovate with clear standards and consumers to have confidence in the products placed on the American market. California has historically enacted policies to promote the safety of products its consumers use, including those with meaningful national security applications, such as drones, Internet of Things devices, and batteries with critical minerals. 

Companies have to reconcile a variety of goals, including a return on investment. Given these foundational financial incentives, policy can play an important role in helping companies stay aligned with valuable goals for the public. However, when regulatory frameworks fail to ensure transparency and accountability, consumer welfare can be significantly compromised.

When industries with incentives to promote particular products possess privileged information about risks while maintaining opacity about their internal research, the resulting information asymmetry undermines effective regulation and public welfare. The history of the tobacco industry demonstrates how, without mandated transparency, corporate interests can systematically distort public understanding, delay appropriate regulatory responses, and compromise public health outcomes for generations.

Robust evidence linking smoking to lung cancer, for example, emerged in the 1950s. Early studies by \citet{dollMortalityDoctorsRelation1950} and \citet{wynderTobaccoSmokingPossible1950}, along with a 1954 American Cancer Society study showing a 52\% higher death rate among smokers \citep{michaelsDoubtTheirProduct2008}, provided quantitative backing for what many doctors had observed anecdotally. Advances in causal inference at the time finally offered the methodological tools to confirm these observations \citep{brandtInventingConflictsInterest2012}, leading to the 1964 U.S. Surgeon General’s Report that formally linked cigarette smoking to lung cancer \citep{schneiderGovernmentalRegulationCigarette1981}. The first regulation of the tobacco industry occurred in 1966 and required companies to include a warning on cigarette packs that ``smoking may be hazardous to your health'' \citep{noarImpactStrengtheningCigarette2016}. By requiring this single line on cigarette packages instead of a more robust warning, regulators failed to provide consumers with broader evidence about the risks that the companies possessed. Companies lobbied for a single line on cigarette packages instead of a more robust warning and actively pushed back against more robust health warnings, including a proposed South Dakota bill that would have required a skull and crossbones on cigarette packages \citep{wagnerCigaretteCountryTobacco1971}. Consequently, larger changes in smoking behavior that could have been achieved with stronger warnings were inhibited.

Internal tobacco industry documents reveal that companies had conducted their own research confirming the harmful effects of smoking, yet they launched sophisticated campaigns to undermine these findings \citep{proctorCancerWarsHow1994}. In 1954, tobacco manufacturers funded public statements emphasizing uncertainty while privately funding research that reinforced the clearly documented relationship between tobacco and health risks \citep{brandtInventingConflictsInterest2012}. One tobacco executive explicitly stated, “Doubt is our product since it is the best means of competing with the ‘body of fact’ that exists in the minds of the general public.” \citep{michaelsDoubtTheirProduct2008} Tobacco companies deliberately shaped public debate by supporting research organizations that would call for more studies, giving the impression of scientific controversy while undermining findings that confirmed smoking’s harmful effects. 

Tobacco companies faced substantial litigation over their lack of transparency from those who had suffered health harms from tobacco. The continued exposure to lawsuits affected the reputation of the companies, as the litigation revealed sophisticated corporate deception efforts. In 1998, the litigation was settled with the Master Settlement Agreement, which was signed by 52 state and territory attorneys general and the country’s four largest tobacco companies \citep{MasterSettlementAgreement1998}. Within the first 25 years after the agreement, tobacco companies were required to make payments estimated to be between \$200 and \$240 billion. Taken together, the tobacco reimbursement lawsuits resulted in the largest redistribution of the costs of corporate wrongdoing in American legal history. Ultimately, companies were coerced to share the evidence they possessed earlier, but only after substantial reputational damage and a decades-long public health crisis.\\

\noindent \textbf{Key themes and lessons:}\\
\textit{Well-calibrated policies can create a thriving entrepreneurial culture for consumer products.} Technological applications have brought, and will continue to bring, enormous benefits to consumers. Thoughtful policy can enhance innovation of these consumer products and promote widespread distribution of their benefits. Carefully designed policies deliver benefits to consumers while creating shared expectations between producers and consumers on the quality and safety of the goods they are consuming, with the potential to create a more robust innovation ecosystem in the long run. Policy must recognize the value of robust markets---including companies with varying sizes and goals---and employ tools that are appropriately designed for those markets. This includes facilitating opportunities
for start-ups and smaller companies with less capital and avoiding burdensome requirements that unfairly support incumbent players. This will be especially true in AI, where scientific and technological development could create demand and potential for extraordinary new applications that could benefit substantial portions of society.\\

\noindent\textit{Transparency is a necessary but insufficient condition for consumers to make informed decisions.} The history of the tobacco industry reveals the importance of developing frameworks that promote transparency around companies’ internal risk assessments and research findings. In the AI context, frontier AI labs possess the most holistic information about their models’ capabilities and risks. Making this information accessible to policymakers and external experts can promote policy informed by a holistic understanding of the state-of-the-art of evidence produced by those closest to the technology, supporting informed oversight without stifling innovation.

Transparency alone is insufficient; as the tobacco case shows, companies can distort public understanding despite available evidence. Independent verification mechanisms are necessary to validate industry claims and ensure that evidence is accurately represented, enabling better policy decisions and preventing potential harms. \\

\noindent\textit{Lack of transparency on product safety can result in avoidable, costly litigation.} 
The tobacco case illustrates that when novel technologies cause injuries and information about safety practices is opaque, litigation is a predictable consequence. In the tobacco case, litigation ultimately brought the requisite information to light, but it resulted in irreversible reputational damage to tobacco companies and devastating public health consequences.

Taken together, the case of consumer products, and especially the value of learning key lessons from society’s experience with products like nonprescription drugs and tobacco, suggest opportunities to push both frontier AI developers and consumers toward the Pareto frontier—that is, carefully targeted regulation can benefit producers and consumers alike. An information-rich environment on safety practices would protect developers from safety-related litigation in cases where their information is made publicly available and, as the next subsection describes, independently verified. Those with suspect safety practices would be most vulnerable to litigation; companies complying with robust safety practices would be able to reduce their exposure to lawsuits.

\subsection{Contending Interests, Accountability, and Useful Evidence From Simulations: Lessons From Energy Policy} 

While consumer products like tobacco most directly affect individual users, the energy industry case demonstrates how information asymmetries can scale to affect entire populations through complex collective action challenges. From the 1960s through the 1980s, energy companies conducted sophisticated climate modeling that accurately predicted global warming trajectories. While energy companies’ internal evidence was conclusive, they publicly and deceptively emphasized uncertainty. The case of fossil fuel companies offers key lessons: Third-party risk assessment could have realigned incentives to reward energy companies innovating responsibly while simultaneously protecting consumers. It also reinforces the importance of relying on broad evidence to build policy that anticipates interconnected technological trajectories, leveraging simulation and modeling techniques that supplement evidence of observed harms. 

By the late 1960s, major energy companies had conducted groundbreaking research about global risks connected to climate change \citep{griffinBriefAmiciCuriae2019}. For example, a Stanford Research Institute report commissioned by the American Petroleum Institute explicitly linked rising levels of carbon dioxide to rising global temperatures and possible consequences including ``melting ice caps, rising sea levels, warming oceans, and serious environmental damage on a global scale.''\citep{robinsonSourcesAbundanceFate1968} Synthesizing a combination of current carbon dioxide levels and expectations around the trajectories of fuel usage, the 1968 report predicted that a 25\% increase in CO2 concentrations by the year 2000 was realistic \citep{robinsonSourcesAbundanceFate1968}. 
 
In the years that followed, energy companies documented these findings while public messaging emphasized uncertainty. For example, researchers at ExxonMobil successfully modeled global warming future trajectories with 63\%–83\% accuracy \citep{supranAssessingExxonMobilsGlobal2023}. These internal company findings aligned with independent academic projections. But in public, ExxonMobil made statements that contradicted their internal findings, framing climate projections as uncertain or speculative. A 1980 policy booklet produced by the American Petroleum Institute (API) titled ``Two Energy Futures: A National Choice for the 80s'' presented an optimistic view of continued fossil fuel expansion while minimizing climate risks. The booklet misleadingly cited prominent scientists and mischaracterized scientific opinions in a way that downplayed concerns \citep{americanpetroleuminstituteTwoEnergyFutures1980}, directly contradicting Carl Sagan’s actual research warning of profound consequences from global temperature changes. This public messaging sharply contrasted with API’s internal knowledge---just six months before the booklet’s publication, their secret industry task force had received briefings from Stanford University engineer John Laurmann warning that climate change could “bring world economic growth to a halt” and require decades of transition away from fossil fuels to prevent catastrophic outcomes \citep{nelsonMinutesAQ9CO21980}. Despite substantial internal evidence that corroborated reports from academia, some industry leaders continued to dismiss the utility of global warming projections. Eventually, like the tobacco industry, energy companies faced lawsuits over their deceit, lack of transparency, and responsibility for damages \citep{supranAssessingExxonMobilsGlobal2023}.\\

\noindent \textbf{Key themes and lessons:}\\
\noindent\textit{Transparency and independent risk assessment are essential to align commercial incentives with public welfare.} 
When industry actors conduct internal research on their technologies’ impacts, a significant information asymmetry can develop between those with privileged access to data and the broader public. As seen in the energy industry case, companies had unique insights into climate impacts that were not fully shared with policymakers or the public. Third-party risk assessment mechanisms could have provided decision-makers with comprehensive evidence needed for more effective policy responses. For frontier AI, establishing independent assessment frameworks that simultaneously account for companies’ intellectual property rights and cross-jurisdictional compliance burdens, potential national security risks of disclosure, and an appropriate degree of transparency into potential public harms can ensure that risk evaluations are not solely controlled by and visible to those with commercial interests in the technology’s advancement. In cases where competitive pressures may disincentivize participation in safety frameworks or unintentionally discourage transparent evidence environments, developing robust policy incentives ensures that developers create and follow through on stated safety practices, such as those articulated in safety frameworks already published by many leading companies. Critically, policy has an important role to play in aligning incentives not just for frontier AI companies but also for third-party auditors, regulatory bodies, and other stakeholders involved in the assessment ecosystem.\\

\noindent\textit{Simulations, modeling, and adversarial testing are useful analytical tools to anticipate future impact.} 
Simulation and modeling are essential tools to anticipate future impacts and help policymakers navigate uncertainties around scientific and technological progress. By projecting potential climate impacts and analyzing relationships between fossil fuel use and environmental change, these tools reveal how individual decisions aggregate into systemic risks. In essence, they expose the collective action problem inherent in managing risk: Only by simulating use at scale, rather than merely testing individual models, can policymakers fully appreciate the cumulative impact of these decisions and design interventions that address both immediate hazards and long-term systemic threats.

In assessing future trajectories for AI, robust simulation techniques and adversarial testing enhance our understanding of risks and potential technological trajectories. Safety evaluations, designed to test AI capabilities in controlled environments through comprehensive red-teaming exercises, provide insights into how systems may behave in possible future, real-world conditions. For example, OpenAI’s o1 system card, building off \citet{meinkeFrontierModelsAre2025}, states, ``When o1 was led to believe that it would be shut down when acting on its goal and that its actions were monitored, it attempted to deactivate this `oversight mechanism' in 5\% of the time'' \citep{OpenAI01SystemCard2024}. Claude 3 Opus, meanwhile, strategically faked alignment in 12\% of the test cases when the model believed its responses would be used for training \citep{greenblattAlignmentFakingLarge2024}. Another report showed o1-preview and DeepSeek R1 attempted to hack their environments as a strategy to win games of chess \citep{bondarenkoDemonstratingSpecificationGaming2025}. Most recently, research from OpenAI showed frontier reasoning models were found to engage in complex reward-hacking behaviors and may learn to obfuscate their intentions when chain-of-thought monitoring is applied during training \citep{bakerMonitoringReasoningModels2025}. These examples collectively demonstrate a concerning pattern: Sophisticated AI systems, when sufficiently capable, may develop deceptive behaviors to achieve their objectives, including circumventing oversight mechanisms designed to ensure their safety \citep{ngoAlignmentProblemDeep2025}. Because these risks are unique to AI compared to other technologies, oversight is critical for external outputs as well as internal testing and safety controls. Policies that govern internal deployment are common for high-risk emerging technologies: In biosecurity, for example, robust regulatory frameworks govern internal research activities regardless of public release status.

While none of these cases likely demonstrate immediate risk, they underscore potential challenges in evaluating the true capabilities of models and our ability to maintain human control over them---including during training, pre-deployment, and deployment inside frontier AI companies \citep{stixAIClosedDoors2025}. Ensuring an appropriate degree of transparency and accountability in evaluating potential long-term consequences and mechanisms will be essential for informed decision-making so that policymakers have access to evidence produced by those closest to technological trajectories. \\
 
\noindent\textit{Comprehensive evidence can highlight key junctures for governance action.} 
Lessons from this case suggest that access to comprehensive evidence allows governments to assess when and how to act on emerging risks, including acting decisively at early stages when the evidence points in that direction. The U.K. government-commissioned Stern Review, which explicitly deployed economic models of uncertainty and risk, estimated that failing to address climate risks could result in the long-term GDP loss of 5\% annually, with wider risks and damages resulting in potential damages up to 20\% \citep{sternSternReviewEconomics2006}. The costs of action to reduce greenhouse gas emissions, meanwhile, were estimated at only 1\% of global GDP each year. This is a useful lesson for AI policy: Leveraging evidence-based projections, even under uncertainty, can reduce long-term economic and security costs. 

\subsection{Brief Examples of Positive Outcomes}

History is also ripe with examples in which evidence-based policy struck an appropriate balance between promoting innovation and ensuring sufficient accountability and transparency, allowing industry to thrive while minimizing risks. In brief, here are three examples:

\subsubsection{Balancing Innovation and Safety and Carrots and Sticks: Lessons From Pesticide Regulation}
The regulatory framework governing pesticide use in the United States exemplifies a balance between safeguarding public health and supporting agricultural productivity.

Governing bodies in this space, the Environmental Protection Agency (EPA) and California’s Department of Toxic Substances Control, for example, evaluate pesticide safety through comprehensive toxicity studies, considering factors such as human exposure, environmental impact, and long-term ecological effects. Once brought under regulation, pesticides are subject to reevaluation and evolving enforcement measures, ensuring that current scientific knowledge informs statutes and regulations.

The United States’ approach to these toxic but necessary substances explicitly accounts for the economic imperatives of the agricultural sector, a cornerstone of national and global food security. California’s agricultural sector alone accounted for \$59.4 billion in receipts in 2023 \citep{californiadepartmentoffoodandagricultureCaliforniaAgriculturalProduction2025}. Pesticides enhance crop yields by mitigating losses due to pests and diseases, thereby stabilizing food supply chains and rural economies \citep{tudiAgricultureDevelopmentPesticide2021}. Regulations, while stringent, offer flexibility through mechanisms such as conditional registrations and exemptions for emergency use, designed to prevent disruptions to critical crop production \citep{usepaPesticideEmergencyExemptions2013}. They also combine both carrots and sticks to incentivize compliance: California’s Department of Pesticide Regulation offers targeted support for sustainable pest management practices \citep{departmentofpesticideregulation$19MillionGrants2024} and levies fines for violations of pesticide use law \citep{DPR23002Civil2024}. Pesticide policy demonstrates the upside of a dual-approach: Targeted support enables high-impact interventions and ensures smaller market participants can adopt improved practices economically; simultaneously, fines offer important deterrents against harmful behaviors that could affect broad segments of society.

\subsubsection{The Importance of Clear Standards: Lessons From Building Codes}
Building codes are constantly evolving to incorporate advances in engineering, green building principles, seismology, and fire resilience, for example. It’s hard to imagine a world where buildings in California did not have standardized stair heights---standardized at 7.5 inches maximum height and 10 inches minimum depth \citep{CaliforniaCodeRegulations2025}. Our brains are conditioned to expect a certain footfall when climbing regular household steps. Similarly, we would not accept a home built to exceed the standard 9-inch spacing for midrails on upper floors, creating a danger for toddlers to slip through \citep{CaliforniaCodeRegulations2025a}. We cannot imagine an unreinforced masonry building being constructed atop a fault line or a house without circuit breakers. 

Despite these rigorous and extremely prescriptive building standards, homes get built every day. The building code---through significant trial and error---has achieved balance between safety and economic output. 

\subsubsection{A False Safety-Innovation Binary: Lessons From Seat Belts}
The National Highway Traffic Safety Administration estimates that in the past four decades, seat belts have saved over 8,900 lives per year \citep{SeatBelts2025}. This is a technology that demonstrably works. But the path to a national requirement was neither fast nor straight.

In 1959, Volvo introduced the first modern, three-point seat belt. It took nearly a decade of fact-finding and debate---and overcoming opposition from the auto industry---for the federal government in the United States to adopt a mandate in 1968 that all new cars include these safety devices \citep{SeatBeltUse2024}. As late as 1984, a Washington Post poll found that 65\% of Americans still opposed mandatory seat belt laws \citep{hendersonNYFirstState1984}. Today, seat belts are an unquestioned, ubiquitous part of daily life. It took decades for this law to go from frontier concept to common acceptance. 

The difference between seat belts and AI are self-evident. The pace of change of AI is many multiples that of cars---while a decades-long debate about seat belts may have been acceptable, society certainly has just a fraction of the time to achieve regulatory clarity on AI.

Although the time scale differs, there are still lessons to be gleaned. The inclusion of seat belts in cars does not impact the main beneficial function of the machine: transporting people from one place to another. Their requirement provides measurable public safety benefits without impeding the utility of the main technology. National seat belt laws have neither made American automakers less competitive than their foreign counterparts nor increased marginal costs to the point of making cars unaffordable.\\

\noindent \textbf{Key themes and lessons:}\\
\noindent\textit{Technology can thrive under a regulated environment as long as policy creates incentives that align company behavior with economic opportunity and public safety, is specific and prescriptive for a particular reason, and does not impair the use of the technology for its stated beneficial purpose.} 
The same can certainly hold true for AI policy in a subnational context such as California. Lessons from previous scientific and technological cases reveal a range of incentive tools available at California’s disposal.\\

\noindent\textit{It is important, however, to acknowledge key distinctions.} 
Pesticides, building codes, and seat belts can be geographically bounded, making policy implementation more straightforward. This is largely true for tobacco as well---the U.S. market can be regulated even with some smuggling. In contrast, computing technologies, climate impacts, and bio-threats transcend geographical boundaries, creating additional governance challenges more similar to those seen in the internet case study.

Additionally, seat belts, tobacco regulations, and, to some extent, building codes represent cases where individual actions provide significant personal risk reduction. This differs fundamentally from AI, where individuals cannot meaningfully “opt out” of societal-level impacts once these systems are widely deployed.

\subsection{Learning the Right Lessons for Future Policy}

In balancing technology’s considerable benefits and risks in the years to come, modern societies have experience to draw from in the development and deployment of technologies that were once labeled advanced and gradually became more integrated with daily life. Although we can learn from and apply those experiences, such knowledge must be used carefully---with due concern for some recurring patterns but also awareness of important distinctions. 

Significant historical experiences with science and technology---inspired by case comparisons to internet policy, consumer products, and energy---highlight the reality that policymaking must often occur under uncertainty. If crafted properly, policymaking under such conditions encourages innovation, pays attention to evidence of both present and emerging risks, and leverages multidisciplinary and thorough assessment of the state of the evidence. Among other implications, the case comparisons underscore the importance of transparency, third-party auditing, and acting on policy windows with mechanisms that surface and enable knowledge and analysis. These insights also highlight the value of augmenting real-world observation with a range of methods to consistently address uncertainty, navigate practical realities, and incorporate critical context from a breadth of perspectives. 

These cases offer certain useful principles to guide governance, including the need to preserve a transparent evidence ecosystem. Nonetheless, the full range of consequences from transformative breakthroughs in AI remain fundamentally uncertain. Along with beneficial consequences in fields such as medicine and science, the future deployment of advanced AI systems could also bring meaningful harms. California has a unique opportunity to shape governance in a way that maximizes the extraordinary benefits AI could unlock while addressing potential risks with foresight and adaptability. The sections that follow offer key principles to inform how California can successfully navigate those trade-offs.

\hypertarget{transparency}{\section{Transparency}}
\label{sec:transparency}
Foundation models are complex technologies driven by several recent breakthroughs in science and engineering: Understanding how these models are built, what they are capable of, and how they are used is essential for effective governance and beneficial for public trust. 
Further, as intermediary technologies in the AI supply chain, foundation models may power applications without users even being aware of it. 
Foundation models may become an invisible force that is profoundly impactful to our society’s functioning. 
By disclosing information to the public, foundation model developers, deployers, and distributors can increase shared understanding of this technology, enabling better decisions. 
However, the most influential foundation model developers are extremely opaque on critical topics. 
The 2024 Foundation Model Transparency Index, which assesses major foundation developers based on developer-submitted reports, shows that developers score on average a 34\% for transparency relating to training data, 31\% for transparency of risk mitigations, and 15\% for transparency of their model’s downstream impact \citep{bommasani2024fmti}. 

Transparency is a fundamental prerequisite of social responsibility and accountability \citep{florini2007right, robinson2012nations, winecoff2024improvinggovernanceoutcomesai}.
Disclosing information that is available, shareable, legible, and verifiable can improve the external visibility into key decisions made by AI companies as well as their impact. 
This includes insight into the motivations for decisions, the evidence to justify the decision, and the cost and benefits that result from it. 
Many organizations regularly call for transparency as a mechanism for fact-finding, accountability, and holding organizations responsible for harm \citep{heikkila2023high, diresta2022openblackbox}.
Shared visibility engenders public trust and facilitates interventions in the public interest \citep{hardin2002trust}. 
Without sufficient understanding of industry practices, the public cannot characterize the societal impact of digital technologies, let alone propose concrete actions to improve business practices \citep{pasquale2015black}. 
While its effects are often difficult to measure as they are diffuse and indirect, transparency helps to expose malpractice and enables the public to respond to such malpractice. 
On the other hand, there are arguments against transparency as well—security and competition, for instance. 
These trade-offs, which are discussed below, are best navigated when transparency is designed to be purposeful, actionable, and outcome-focused.

In addition to benefiting many stakeholders throughout the AI ecosystem and the broader public, increased transparency around foundation models is uniquely beneficial to the government for three reasons. 
First, when the government procures AI technology, whether foundation models directly or services built on top of them, information is required to ensure due diligence in the government’s use of AI \citep{omb2024ai, calgovops2024ai}.
Second, when the government designs policy, a clear understanding of the developer practices will support better designed policy. For example, transparency into the safety practices of developers helps clarify if further policy interventions are required to lower risk to acceptable levels. 
Finally, when the government enforces regulation, a clear understanding of a developer’s conduct, especially in addressing safety concerns, will enable better-targeted application of limited enforcement resources.

Several existing policies recommend or mandate the improved transparency surrounding foundation models. During the 2024 legislative session, the California legislature passed and Governor Gavin Newsom signed into law AB 2013, which requires public-facing transparency into the training data used to build generative AI systems. 
The EU AI Act requires similar public-facing transparency into training data and copyright policies along with further disclosures on a range of topics, including risk mitigation, directed at the EU and national governments as well as downstream firms in the AI supply chain. 
Voluntary commitments secured by the Biden administration and the G7 indicate that companies should release public-facing transparency reports \citep{bommasani2024foundationmodeltransparencyreports}, and the Frontier AI Safety Commitments signed at the AI Seoul Summit 2024 reinforce the focus on public transparency with further detail to be provided to trusted actors, including developers’ home governments. 
However, these existing approaches adopted within and beyond California do not fully address crucial elements of transparency.
With that said, any future policy developed on transparency in California should proactively consider the overlap, and harmonize where appropriate, with preexisting policy (possibly in other jurisdictions) to minimize duplicative compliance burdens for companies operating in multiple jurisdictions.
We highlight below how policymakers may strengthen public-facing transparency in key areas.

Transparency is ultimately an instrumental, rather than terminal, goal for policy: It provides the necessary foundation for accountability, but it is often insufficient and requires supplementary verification mechanisms to provide the accountability benefits in full. 
In some cases, transparency to a carefully scoped set of actors, such as a government agency or certified, credible, third-party auditor, may be appropriate, depending on policy intent.

A ``trust but verify'' approach recognizes the distinctive insights and expertise of foundation model developers about their technology while ensuring claims can be independently validated. 
This second step, which is reinforced by whistleblower protections and third-party research protections, engenders public confidence to enable evidence-based governance.

\subsection{Improving Transparency}
As policymakers explore how to increase transparency, two core issues emerge: (i) what information should be disclosed; and (ii) to whom it should be disclosed \citep{bommasani2024foundationmodeltransparencyreports, kolt2024responsiblereportingfrontierai}.
To address the former, we identify key areas where additional transparency would be desirable. 
These include areas where there is systemic opacity despite clear arguments for why greater transparency would be valuable. As a key resource, we leverage the 2024 Foundation Model Transparency Index, which holistically scores 14 major foundation model developers (\eg, Anthropic, Google, Meta, Mistral, OpenAI) for their level of public-facing transparency as of 2024 \citep{bommasani2024fmti}.
To address the latter, transparency measures should consider who is best positioned to use information. 
Since citizens, journalists, academics, and civil society are often the first to identify technological harms, especially given limited governmental resources, policy should often prioritize public-facing transparency to best advance accountability. \\

\noindent \textit{Data acquisition.} 
Transparency into how data is obtained (\eg, via crawling the internet, licensing from data vendors, paying for human labor) bears on competition (\eg, if data is exclusively licensed to one model developer), safety (\eg, if child sexual abuse material is in the data), security (\eg, if data is able to be poisoned), and copyright (\eg, if data was acquired with the consent of its owners). In spite of these compelling reasons for transparency, the 2024 Foundation Model Transparency Index documents an average score of 32\% across major foundation model developers for data sourcing transparency. While California passed AB 2013 on data transparency in 2024, and the EU AI Act mandates a public-facing summary of training data for all general-purpose AI models on the EU market, more dedicated visibility into data acquisition would positively address key interests. \\

\noindent \textit{Safety practices.}
Transparency into the risks associated with foundation models, what mitigations are implemented to address risks, and how the two interrelate is the foundation for understanding how model developers manage risk. 
In turn, this information directly informs how other entities in the supply chain should modify or implement safety practices.
In addition, transparency into the safety cases used to assess risk provides clarity into how developers justify decisions around model safety \citep{hilton2025safetycasesscalableapproach}.
The 2024 Foundation Model Transparency Index documents average scores of 47\% and 31\% for risk-related and mitigation-related transparency, respectively. 
California’s proposed SB 1047, which was vetoed by Governor Newsom in 2024, in part aimed to provide transparency into safety practices as do the voluntary Frontier AI Safety Commitments and mandatory EU AI Act obligations. \\

\noindent \textit{Security practices.} Transparency into the cybersecurity practices of model developers, namely their ability to secure unreleased model weights from both company-internal and company-external exfiltration, is a clear priority given that model weights are high-value assets with broad-ranging societal dependence in many downstream AI applications. Prior work \citep{nevo2024securing} extensively catalogues how developers can improve cybersecurity practices for model weights, identifying several practices where it is currently unclear if leading developers do or do not implement these practices (\eg, insider threat detection, access controls, confidential computing). 
Weights-level cybersecurity has emerged as a clear global priority, including in the G7 International Code of Conduct and the Frontier AI Safety Commitments. \\

\noindent \textit{Pre-deployment testing.} Transparency into pre-deployment assessments of capabilities and risks, spanning both developer-conducted and externally conducted evaluations, is vital given that these evaluations are early indicators of how models may affect society and may be interpreted (potentially undesirably) as safety assurances. Specifically related to external evaluations, clarity into the level of rigor (\eg, the time and depth of pre-deployment access provided) and independence (\eg, if the external party is paid by the model developer and whether they are constrained in what they can disclose) is crucial for ensuring the public understanding is calibrated in interpreting these pre-deployment evaluations. Recent incidents, such as the pre-deployment assessment of model capabilities of OpenAI’s o3 by Epoch on the FrontierMath benchmark, indicate the importance of concretizing these norms for assessments in critical safety areas. \\

\noindent \textit{Downstream impact.} 
Transparency into how foundation models are deployed across the economy is a core prerequisite for measuring the impact of AI in society, projecting how AI will affect the future of work, identifying where AI demonstrably causes harm, and grounding policymaking \citep{bommasani2023ecosystem}. 
Namely, while foundation models are general-purpose technologies \citep{bresnahan1995gpt, bommasani2021opportunities, eloundou2023gpts} that can, in theory, be used in almost every domain, they will in practice have a more particular usage distribution. 
For example, Anthropic’s Economic Index shows Claude usage is concentrated on software development and technical writing tasks \citep{tamkin2024clio, handa2025economictasksperformedai}.
To monitor downstream impact, attention should be allocated to entities that host models for download (to track the number of downloads) or for inference (to track the number of queries or users): For example, these entities could publish periodic reports on the adoption of foundation models throughout the economy.
Critically, by not necessarily focusing on model developers but instead on distribution channels, such an approach can help track the impact of open models even though their developers may be ill-equipped to directly track impact \citep{kapoor2024position}. 
The 2024 Foundation Model Transparency Index finds that downstream impact is the area of systemically worst transparency with developers, scoring just 15\% on average. 
Alternative approaches include market surveillance by government agencies, as the U.K. Competition and Markets Authority has begun to explore \citep{cma2023ai}. \\

\noindent \textit{Accounting for openness}.
Foundation models are released using a variety of approaches
\citep{solaiman2019release, liang2022community-norms, shevlane2022structured, solaiman2023gradient}, which shape how they can be accessed by entities beyond their developers \citep{solaiman2025release}.
The most salient distinction is based on whether the model weights are made broadly available (\ie, open-weight models like Llama 3.3 and DeepSeek-R1) or not (\ie, closed or proprietary models like Gemini 2.0 and Claude 3.7).
Release strategies generate challenging trade-offs between the benefits and risks of models: Open models are often associated with benefits to innovation, competition, privacy, and science, but they may be more prime for misuse by malicious actors \citep{kapoor2024position, ntia2024open}.
Openness and transparency are not the same, and often are conflated with each other, but they are related.
Open model weights enable deeper inspection, which is necessary for several forms of scientific research, including on safety and security \citep{kapoor2024position, bommasani2024open}. 
Beyond the direct benefits to transparency afforded by openness, open model developers have so far demonstrated greater transparency, though heightened transparency is not guaranteed by a developer’s decision to make models open. 
The 2024 Foundation Model Transparency Index demonstrates a correlation: Open-weight model developers are more transparent on average, especially in terms of how their models are built.
As a result of the benefits of open models, including on the topic of transparency, the EU AI Act grants an exemption for open models from the transparency obligations under the AI Act, except where models pose systemic risk.

\subsection{Improving Third-Party Risk Assessment}
Greater transparency is integral to an improved AI governance ecosystem; however, despite its considerable value, it is not sufficient without a broader ecosystem to complement it. 
Fundamentally, this is because the value of transparency from model developers is naturally limited by the information these developers have. 
For a nascent and complex technology being developed and adopted at a remarkably swift pace, developers alone are simply inadequate at fully understanding the technology and, especially, its risks and harms.

Third-party risk assessment, therefore, is essential for building a more complete evidence base on the risks of foundation models and creating incentives for developers to increase the safety of their models. 
At present, some prominent foundation models are subject to risk evaluations by first-party model developers or second-party contractors. 
In contrast, third-party evaluation affords three distinctive strengths. 
First, third-party evaluations have unmatched scale: Thousands of individuals are willing to engage in risk evaluation, dwarfing the scale of internal or contracted teams. 
Second, third-party evaluations have unmatched diversity, especially when developers primarily reflect certain demographics and geographies that are often very different from those most adversely impacted by AI. 
Broad demographic, institutional, and disciplinary diversity is vital for unearthing blind spots. 
And finally, third-party evaluation is distinctively independent: Society requires forthright and trustworthy assessments of risk, which benefits from a lack of commercial and contractual entanglement with AI developers. 

A well-designed disclosure and verification regime offers significant benefits not only to the public but also to foundation model developers. 
By establishing industry-wide transparency standards and third-party verification mechanisms, companies can demonstrate compliance with best practices, potentially reducing their liability exposure compared to the uncertainties of purely reactive litigation. In practice, states could leverage the existing technical infrastructure and expertise of federal-level AI institutions to conduct standardized risk evaluations.

Furthermore, this approach can transform competitive dynamics around safety. When safety measures and risk assessments are publicly disclosed and verified, companies face stronger incentives to implement best practices, as deviations would be apparent and could attract greater scrutiny. 
This transparency coupled with third-party verification effectively creates a ``race to the top'' rather than a ``race to the bottom'' in safety practices, benefiting responsible companies while improving overall industry standards. \\

\noindent \textit{Researcher protections.} 
Third-party AI evaluators require access to prominent foundation models and permission from developers to conduct risk assessments. While most major models are available in some form to the public, whether via an API (\eg OpenAI’s GPT-4o) or via their weights (\eg Meta’s Llama 3.3), the key challenge is whether evaluators are permitted to conduct risk evaluations. 
Prior work documents that ``companies disincentivize safety research by implicitly threatening to ban independent researchers that demonstrate safety flaws in their systems.'' \citep{klyman2024fas} 
These suppressive effects come about due to companies’ terms of service, which often legally prohibit AI safety and trustworthiness research, in effect threatening anyone who conducts such research with bans from their platforms or even legal action \citep{klyman2024acceptableusepoliciesfoundation}. 
In response, in March 2024, over 350 leading AI researchers and advocates signed an open letter calling for a safe harbor for independent AI evaluation.\footnote{\url{https://sites.mit.edu/ai-safe-harbor/}} 
Such a safe harbor, analogous to those afforded to third-party cybersecurity testers, would indemnify public interest safety research and encourage the growth of the evidence base on AI risk \citep{longpre2024safeharboraievaluation}.\footnote{To avoid confusion, we emphasize that the safe harbor described in this section affords protections to third-party researchers. This is distinct from proposals that may afford protections to other entities, such as foundation model developers.} \\

\noindent \textit{Responsible disclosure. }
Beyond establishing legal protections for evaluators, responsible dissemination of findings presents another critical challenge. Developing norms around responsible risk evaluation requires the definition of rules of engagement: Which foundation models or AI systems should be assessed, and what prior notice should be given to their developers and providers in advance of broader dissemination of vulnerabilities? 
Beyond improving the practices of third-party evaluators, infrastructure is sorely needed to communicate identified vulnerabilities to the many affected parties. First, risks identified at the foundation model level may implicate or impact parties elsewhere in the supply chain. 
For example, if a text-to-image foundation model readily generates synthetic child sexual abuse imagery, upstream data sources should inspect whether real child sexual abuse images are present in their datasets, and downstream AI applications should also inspect whether these applications are susceptible to generating similar imagery \citep{thiel2023csam, thiel2023generative}.
Second, risks at the foundation model level are often transferable: Many foundation models, potentially created by different developers, may share the same vulnerability. For example, prior work has shown that jailbreaks of one language model can be reused to jailbreak many other language models \citep{zou2023universal}.
For these reasons, routing mechanisms are needed to communicate third-party risk assessments to not just the foundation model developer but also the other relevant parties who may need to take actions to reduce risk. 
Infrastructure used to route third-party risk assessments could likely be reused to share information on adverse events as we describe in Section 4.

\subsection{Protecting Whistleblowers}
Most information that is shared about the practices of foundation model developers is deliberately disclosed by the developer, whether due to a voluntary choice, industry standard, or mandatory requirement. However, under exceptional circumstances, employees of the developer or other parties may disclose information. In this vein, in 2024, several current and former employees at frontier AI companies, as well as AI researchers, signed an open letter arguing for ``A Right to Warn about Advanced Artificial Intelligence.''\footnote{\url{https://righttowarn.ai/}}

Whistleblower protections are defined as legal safeguards that prohibit retaliation against individuals (generally employees) who report misconduct, illegal activities, fraud, or other wrongdoing within an organization. ``Whistleblowers can play a critical role in surfacing misconduct, identifying systemic risks, and fostering accountability in AI development and deployment.'' \citep{wu2025aiwhistleblowing} Based on the Whistleblower Protection Act of 1989, we frame the design space for essential elements of specific whistleblower protections in relation to foundation model developers and frontier AI \citep{whitaker2007whistleblower}.\footnote{We deliberately avoid exploring the enforcement mechanisms and the interplay with other contractual obligations such as non-disclosure agreements \citep{short1998killing}, which are essential considerations in designing whistleblower policy.} \\

\noindent \textit{Who is eligible for coverage?}
Most existing whistleblower protections apply to all employees within an organization, especially in relation to private sector employees. 
In some cases, differentiation may also relate to coverage under a different whistleblower protection: The Whistleblower Protection Act applies to many federal employees, while military personnel are not covered due to protection afforded them by the Military Whistleblower Protection Act, and employees of intelligence agencies are not covered due to protections like the Intelligence Community Whistleblower Protection Act. 
However, a central question in the AI context is whether protections apply to additional parties, such as contractors. 
Broader coverage may provide stronger accountability benefits but also imposes greater cost: To extend protections to contractors and third parties, developers may need to implement additional reporting channels and legal frameworks.\\ 

\noindent \textit{Under what conditions are disclosures eligible for coverage?}
Different existing whistleblower protections tend to apply when two conditions are satisfied: (i) The whistleblower is blowing the whistle on appropriate topics; and (ii) the whistleblower follows established reporting protocols. In terms of the topics that qualify for protection, prior work \citep{nie2025whistleblower}, based on a survey of existing whistleblower protections across multiple jurisdictions (\eg, the United States at the federal level, the European Union), finds that many existing protections across different sectors share a focus on violations of the law. However, actions that may clearly pose a risk and violate company policies (\eg, releasing a model without following the protocol laid out in a company’s safety policy) may not violate any existing laws. 
Therefore, policymakers may consider protections that cover a broader range of activities, which may draw upon notions of ``good faith'' reporting on risks found in other domains such as cybersecurity \citep{longpre2024safeharboraievaluation}.

In terms of the reporting protocol, disclosures may be directed at different parties (\eg, a company’s board or a designated entity in the government) and have established reporting channels to facilitate the secure (and potentially anonymous) communication of information. The European Union’s AI Act includes explicit provisions on whistleblower protections, which build on the EU Whistleblower Directive and provide the ``Right to Lodge a Complaint with a Market Surveillance Authority'' in the case of AI Act infringement. Ensuring clarity on the process for whistleblowers to safely report information can jointly advance accountability and manage countervailing interests, such as the disclosure of trade secrets or the misuse of information to compromise safety and security. \\

\noindent \textit{Applicability of existing whistleblower protections.} 
Consistent with California Executive Order N-12-23, existing regulatory authority can be considered for AI-related whistleblowing. In general, existing whistleblower protections are stronger for public sector employees than private sector employees \citep{nie2025whistleblower, wu2025aiwhistleblowing}, a fact worth considering given that most foundation models are built by private sector companies \citep{maslej2024artificialintelligenceindexreport}. In some cases, public sector whistleblower protections may apply to private sector employees when the government procures AI from the private sector \citep{nie2025whistleblower}.

\subsection{Guiding Principles}
Many fundamental values and interests are often presented as being at odds with transparency. 
These include privacy (transparency may reveal personal information), security (transparency may reveal vulnerabilities), intellectual property and trade secrets (transparency may compromise competitive advantage), expression (reporting obligations may compel speech), and innovation (reporting obligations may delay progress). 
While policy in many domains actively grapples with the very real trade-offs between these values, we emphasize that foundation models are an immature technology with immature norms. 
In other words, we strongly believe that individual companies and the industry as a whole is not at the Pareto frontier. 
This means that improvements are possible to one or both values at this moment without immediately confronting a trade-off. 
In turn, strategic approaches can increase transparency on foundation models and their supply chain at minimal, if any, cost to these other key priorities. 

Alongside concerns that transparency is harmful or bureaucratic, a separate class of concerns stems from the belief that transparency is not enough. 
Most sharply, transparency in itself does not guarantee accountability or redress: If AI technologies do mediate harm, transparency alone will not rectify these harms, nor will it inoculate us from future harms. 
Therefore, by focusing on transparency, some scholars argue this will advance ``transparency washing'' \citep{zalnieriute2021transparency}, meaning the futile pursuit of transparency without achieving more substantive change or improvement. 

To abate each of these concerns, we identify matching principles: Increasing transparency, subject to the described principle, will yield Pareto improvements.\\

\noindent \textit{Privacy.} 
In the context of improving transparency for foundation models and their supply chain, the vast majority of the desired information does not involve disclosing information that would undermine privacy, such as personally identifiable information. Therefore, privacy is overwhelmingly not an issue for the disclosures we discuss, though privacy could be an issue if disclosures address company-internal training data or detailed usage data. \\

\noindent \textit{Security.} 
General details about risks of foundation models can be made public without undermining security, especially if these risks have been demonstrated in other foundation models or AI technologies. Specific details about concrete vulnerabilities should be disclosed carefully, with advanced notice to actors in the supply chain who are able to remediate them prior to broader disclosure. Computer security, a closely related area, strongly suggests that ``security through obscurity'' will not work for AI, and that responsible, coordinated flaw disclosure can make systems and users safer. \\

\noindent \textit{Intellectual property and trade secrets.} 
Information that is known by a developer’s competitors can be disclosed publicly to provide public benefit with minimal, if any, cost to competitive advantage. Sensitive commercial information that is difficult to make public may also be disclosed to the government at little risk, assuming the government can provide assurance of good information security practices. \\

\noindent \textit{Freedom of expression.}
In the United States, the First Amendment's free speech clause prohibits the government from abridging the freedom of speech.
While not prohibiting speech, disclosure requirements may introduce a concern of First Amendment violation by compelling speech \citep{crs2023first}.
To avoid these concerns, legal precedent indicates that mandates are more likely to be upheld as constitutional if the information is ``purely factual and uncontroversial'' rather than ``highly subjective''  and the disclosure is not
``unduly burdensome'' \cite{cdt2025first}. \\

\noindent \textit{Innovation.} 
Much of the information that would be disclosed as part of a new transparency regime is already produced by companies for their internal operations. Sharing this information, therefore, is considerably less costly because companies already possess it, and the government can decrease reporting cost by creating an effective reporting interface, minimizing duplicative or redundant reporting, and prioritizing directly actionable information. In addition, reduced overall litigation exposure through ``race to the top'' market dynamics will reduce holistic developer costs. \\

\noindent \textit{Transparency washing.} Public-facing transparency makes it much more likely that information disclosure will advance accountability through the joint efforts of governments, media, civil society, and academic researchers, leveraging clear standards to facilitate accountability. A concrete plan of action for how to analyze and base future decisions on disclosed information can also ensure transparency results in tangible progress. 
\hypertarget{aer}{\section{Adverse Event Reporting}}
\label{sec:aer}
Effective AI governance requires a grounded understanding of how AI technology mediates impact in society. 
To better understand the practical impact and, in particular, the incidents or harms associated with AI, we need to develop better post-deployment monitoring \citep{stein2024rolegovernmentsincreasinginterconnected}.
In other domains, post-deployment monitoring by the government yields reports of medical complications, equipment malfunctions, near misses, and other unforeseen hazards. 
By aggregating information about incidents from developers and users, an adverse event reporting system that allows developers and downstream users to report post-deployment incidents promotes continuous learning \citep{johnson2003handbook}, reduces information asymmetries between government and industry \citep{naiac2023aers}, and enables both regulators and industry to coordinate on mitigation efforts \citep{williams2018safety}. 
As a result, an adverse event reporting system constitutes a critical first step toward data-driven assessments of the benefits and costs of regulatory actions \citep{guha2023ai}.  

Adverse event reporting\footnote{The term ``adverse event reporting'' is often applied in the context of medical or public health monitoring systems (\eg, VAERS, FAERS, MedWatch). In other domains, such as in transportation or occupational safety monitoring (\eg, CISA, NTSB safety, accident, or hazard reports), the term ``incident reporting'' is used. Irrespective of the name, these systems monitor for unexpected risks that can lead to early mitigation efforts. } 
refers to a proactive monitoring system designed to collect information about relevant events or incidents from various mandated or voluntary reporters. 
Adverse event reporting systems administered by the government aim to improve access to information about harms to enable regulators, as well as industry actors and users, to learn about risks. 
The systems provide a starting point for regulation without necessarily subjecting technology, firms, or other entities to additional requirements. 
Rather, the goal of these regimes is to aggregate information that may be used to warn users about risks \citep{nhtsa_safety_recall_data}, inform future regulatory efforts, or develop best practices. 

Implementing an adverse event reporting system requires specifying (i) the events to be reported, (ii) the information to report about these events, (iii) the entities that are mandated to report adverse events and those that are permitted to report such events, and (iv) the entities that receive information about individual events or aggregated statistics \citep{cset_components_reporting, guha2023ai}. 

\begin{itemize}

\item \textbf{Reportable events.} 
Adverse event reporting systems designate conditions under which an event should be reported, defined in terms of realized harm or other unintended/detrimental outcomes. \\

\item \textbf{Reported information.} 
The system specifies what information must be shared and how it should be formatted. To appropriately incentivize reporting, some systems permit anonymized or confidential reporting. For example, the Aviation Safety Reporting System (ASRS), administered by the National Aeronautics and Space Administration (NASA) on behalf of the Federal Aviation Administration (FAA), collects confidential reports on aviation safety incidents \citep{nasa_asrs}. \\

\item \textbf{Reporting entities.} 
In different domains, manufacturers, companies, practitioners, users, or any impacted party may be required or permitted to report. 
Typically, mandatory reporting requirements apply to business entities, whereas users or the public may be permitted, but not required, to report. 
Weighing the reliability of reports against the resources required to process submissions is critical \citep{getz2014evaluating}. \\

\item \textbf{Information sharing.} 
Information on commonly reported adverse events may be summarized and shared with enforcement agencies, non-enforcement agencies, industry actors, upstream actors, downstream actors, or the public.
\citep{faers_dashboard}. 
For example, the U.S. Cybersecurity and Information Security Agency (CISA) has implemented several information sharing programs, including the development of Automated Indicator Sharing (AIS) and coordinated vulnerability disclosure process, and participation in the Joint Cyber Defense Collaborative (JCDC) and Joint Ransomware Task Force (JRTF). 

\end{itemize}

Several policy domains implement adverse event reporting systems.
In transportation safety, aircraft incidents are reported to the NTSB and vehicle safety defects are reported to the NHTSA.\footnote{The NTSB administers reporting systems for aviation incidents, which include limited enforcement immunity to encourage voluntary reporting of safety issues without fear of punishment. The NHTSA administers an incident reporting portal for defects.}
In public health, post-vaccination adverse events are reported to the Vaccine Adverse Event Reporting System jointly administered by the CDC and FDA.\footnote{The Center for Disease Control and Prevention (CDC) and the Food and Drug Administration (FDA) jointly administer the Vaccine Adverse Event Reporting System (VAERS) to collect and analyze reports of adverse events (potential side effects or reactions) that occur after vaccination.} 
In cybersecurity, cyber incidents and vulnerabilities are reported via programs administered by CISA \citep{cisa_vulnerabilities}.\footnote{CISA administers incident reporting under the Cyber Incident Reporting for Critical Infrastructure Act of 2022 (CIRCIA) and also administers a Vulnerability Disclosure Policy (VDP) platform which enables researchers to report vulnerabilities to participating agencies.} 

In the domain of AI, the Organisation for Economic Co-operation and Development (OECD) recently developed an incident reporting tool called the AI Incidents Monitor (AIM), which provides a platform for stakeholders to see information about where AI incidents have been reported globally. 
AIM is powered by a machine learning model that monitors news reports about AI incidents and compiles information about the magnitude of their impacts. 
While it is completely voluntary and relies mostly on information reported in the press, it is an example of how to design a system to reach a broad audience.
As additional efforts emerge to document post-deployment impacts, including adverse event, efforts should be made to consolidate reports to avoid reporting fatigue due to multiple fragmented processes. 

\subsection{Benefits} 
We identify four benefits that would arise from well-implemented adverse event reporting in AI, building upon the National Artificial Intelligence Advisory Committee (NAIAC) recommendation to ``Improve Monitoring of Emerging Risks from AI through Adverse Event Reporting'' \citep{naiac2023aers}.\footnote{A subset of the authors of this report also authored the NAIAC recommendation.}\\

\begin{enumerate}
\item \textbf {Improved identification of near-term, potentially unanticipated, sociotechnical harms.}
An adverse event reporting system is able to identify new sources of risk without requiring regulators to specify the harms being monitored for \textit{ex ante}.\footnote{We note that an adverse event reporting scheme for AI is not a substitute for other post-deployment risk assessments, like randomized control trials or large-scale studies, but rather complements these modes of assessment.} 
By defining incidents in terms of unwanted, dangerous, or damaging output \citep{hhs_unanticipated_problems}, such as events involving material harms to life, physical property, or financial interests, regulators learn about threats that materialize without having to enumerate every possible harm or source of risk. 
The reporting system generates data to allow regulators to perform independent risk assessments.\\

\item \textbf{Potential prevention of costly accidents by encouraging proactive mitigation.}
Though incident reporting systems may be costly to establish and maintain \citep{CISA_projected_cost}, there is evidence that they can prevent costly incidents or recalls. 
Moreover, reporting of widespread or critical issues may lead entities to take proactive, voluntary measures to mitigate risk. 
In particular, organizations may identify issues that, if left unaddressed, could lead to systemic failures, reputational damage, or financial liability.
Additionally, the aggregated data can highlight latent vulnerabilities shared across multiple systems, allowing industry actors to collaboratively develop more robust design or operational practices \citep{wansley2023regulating}. \\

\item \textbf{Improved coordination between the government and the private sector in risk mitigation.} 
A major benefit of an adverse event reporting system is the ability to transmit relevant information to relevant actors quickly. 
This may encourage industry to adopt standardized reporting, data collection, or mitigation procedures. 
By standardizing reporting and data collection, the system enables regulators and industry stakeholders to detect patterns, assess systemic risks, and develop targeted responses more effectively. 
Structured reporting mechanisms further help to clarify compliance expectations, streamline information-sharing, and enhance regulatory oversight while reducing administrative burdens for involved parties. 
Over time, this coordination strengthens AI governance efforts by ensuring that both regulators and private entities operate with a shared understanding of risks which, in turn, can enable more effective and proactive responses to emerging threats.\\

\item \textbf{Limited cost imposed on reporting entities and the government. }
Compared with other regulatory proposals to measure and reduce risks associated with AI, adverse event reporting may be comparatively less costly for government agencies administering the system as well as for reporting entities. Reporting entities only experience a reporting cost when an adverse event occurs, and reporting requirements can be targeted, continuously adjusted, and graduated—with the amount of required supporting documentation calibrated to the severity of the incident. Moreover, streamlining reporting forms, allowing electronic submissions through an online platform, clarifying requirements, and allowing summarization \citep{aer_cancer_trials} can reduce the burden of reporting obligations for organizations and individuals. Government agencies may face lower operational costs due to the passive nature of the reporting system, which primarily relies on entities to submit reports rather than requiring continuous monitoring or direct oversight. The use of standardized reporting mechanisms and automated data processing can further reduce administrative costs by minimizing manual review and facilitating efficient data analysis. Additionally, agencies can leverage existing regulatory infrastructure, such as integrating the reporting system into current oversight frameworks, reducing the need for new institutional structures or extensive resource allocation.

\end{enumerate}

\subsection{Challenges}
We identify three general challenges for the successful implementation of adverse event reporting, which we further contextualize to address the distinctive characteristics of AI.

\begin{enumerate}

\item \textbf{Defining clear criteria for what events should be reported.}\\
First and foremost, an adverse event reporting system must provide a clear definition of reportable events. Because the capabilities and risks of frontier models are to a large extent uncertain, the reporting criteria should be broad enough to encompass unforeseen or emergent sources of risk, but sufficiently tailored to avoid complicating risk assessments \citep{cisa_comment_on_rule}. 
In other settings, adverse events are defined in terms of unanticipated or serious health outcomes, financial consequences, or device malfunction. \\ 

\item \textbf{Creating institutional capacity to monitor and investigate adverse events.}\\ 
An effective adverse event reporting system requires regulators to continuously monitor and assess reports of harms, which requires an ongoing commitment of resources and personnel to manage. Analyzing and investigating reports may be costly. 
For instance, the estimated cost for CISA to administer and review mandatory cyber-incident reports over a 10-year period is estimated to be in the billions of dollars \citep{CISA_projected_cost}. 
However, the benefits, including the rapid identification of cross-system vulnerabilities and secure development \citep{CISA_projected_cost}, likely far exceed the operational costs given the wide-ranging estimates of the cost of cyber incidents. 
Moreover, clearly defined reporting criteria and information to be submitted can reduce the burden of investigating reports by ensuring regulators have access to key information about incidents. \\

\item \textbf{Equipped and incentivized to address the underreporting of events and other non-compliance with the system.} \\
Some evidence suggests low rates of compliance with even mandatory adverse event reporting requirements in other contexts, such as adverse vaccine reactions \citep{bailey2016adverse}. 
As a result, adverse event reporting systems may be more effective at monitoring for common interest risks, for which reporters have a shared interest in reporting adverse events, or when combined with safe harbors or other incentives to report \citep{gao2021dynamics, nasa_asrs}. 
Furthermore, in order for reports to be accurate and informative, entities must be equipped to monitor for the reportable event.  

\end{enumerate}

\subsection{Guiding Principles}
Adverse event reporting systems address a core impediment to targeted AI regulation by enabling regulators to learn about realized harms and unanticipated sources of risk. 
Adverse event reporting systems function well when monitoring for risks that regulators and industry have a common interest in addressing, when reports are shared with the appropriate or relevant actors, when agencies have the staff and resources to review reports, when reporting requirements are easily modified, when reporting interfaces are user-friendly \citep{craigle2007medwatch}, and when agencies leverage existing regulatory authority in early mitigation efforts.
Based on our review of similar efforts for cybersecurity threats and AI incidents, including recurring challenges, we offer four targeted principles that can inform a carefully crafted, government-administered adverse event reporting system for AI. \\

\noindent \textit{Sharing adverse event reports with agencies with relevant domain-specific authority and expertise is crucial to collaboratively address identified harms.} Providing access to adverse event reports to relevant agencies with domain-specific regulatory authority and expertise can address the information asymmetry between government and developers in terms of where different risks are occurring. Reports can be made confidential for the government to protect proprietary information within closed systems. However, it could be worth giving agencies the authority to share anonymized findings from reports with other industry stakeholders if an agency determines that other developers could benefit from a finding. In high-risk domains, like those with national security implications, information-sharing practices are common—for instance, the joint ventures, like the Joint Cyber Defense Collaborative (JCDC) and Information Sharing and Analysis Organizations (ISAO), and procedures, like the coordinated vulnerability disclosure process, facilitate access to threat indicators. \\

\noindent \textit{An initially narrow adverse event reporting criteria built around a tightly defined set of harms promotes consistency and comprehensiveness in reporting.} 
Thereafter, periodic review and analysis in aggregate can inform updates to reporting criteria.
Defining adverse events narrowly initially ensures reports are consistent and comprehensive: Past efforts in other domains have sometimes suffered from overly inclusive reporting criteria, which risk drowning signal in noise.
In addition, we recommend that agencies responsible for managing an adverse event reporting system frequently review submissions to identify unanticipated or new sources of risk and to evaluate the need to revise reporting criteria in response to data received through any such reporting system. 
Ideally, adverse event reporting is an iterative learning process \citep{ca_2016_databreach}.\footnote{California’s Data Breach Notification Law, Cal. Civ. Code \S~1798.82, has been amended several times to expand the scope of covered information based on evidence of new threats.}
Information shared on incidents helps to enhance our understanding of how systems are functioning, but a need will remain for agencies to constantly and proactively look for new gaps in information and to identify how reporting systems can help fill those gaps. Policymakers could also consider granting agencies sufficient authority to expand the scope of reporting criteria as appropriate and adopting reporting criteria tailored to California companies. \\

\noindent \textit{An adverse event reporting system that combines mandatory developer reporting with voluntary user reporting maximally grows the evidence base.} A hybrid model of mandatory and voluntary reporting requirements in designing an adverse event reporting system can maximize the robust evidence base necessary for adverse event reporting systems to function properly \citep{cset_hybrid_reporting, cset_ai_incidents}. 
For example, a system could require mandatory reporting for AI model developers that operates in tandem with voluntary reporting for downstream users.

An effective hybrid approach may place mandatory requirements on parts of the AI stack where information gathering is most critical while structuring the right incentives to encourage voluntary reporting in other places. Another component of this approach could be to incentivize manufacturers to engage in verifiable pre-market testing as a way to satisfy certain reporting requirements. 
For example, if a developer engages in sufficient pre-market testing and releases a summary of its results, it could then be exempted from the mandatory requirements.  

Hybrid reporting systems have been used successfully to monitor for post-deployment risks. For instance, cybersecurity incident reporting systems administered by CISA follow a hybrid model: The web-based portal accepts both voluntary and mandatory reports. \\

\noindent \textit{Non-AI-specific regulatory authorities could be used to mitigate the risks of AI technology.} Existing regulatory authority could offer a clear pathway to address risks uncovered by an adverse event reporting system, consistent with California Executive Order N-12-23. This is critical as the rapid pace of AI advancements may challenge regulators’ abilities to craft regulations targeted at specific risks. To address risks and harms of AI does not necessarily require AI-specific regulatory authority or tools. In addition, reviewing existing regulatory authorities can help agencies identify regulatory gaps—in particular, what new authorities are necessary to address risks. 
\hypertarget{scoping}{\section{Scoping}}
\label{sec:scoping}
Well-designed regulation is proportionate: The obligations government imposes on an actor are commensurate with the associated risk to ensure regulatory burdens are such that organizations have the resources to comply and not so burdensome as to curb innovation. The United States, along with 29 other governments, committed to adequately considering a proportionate governance approach by signing the Bletchley Declaration at the 2023 U.K. AI Safety Summit: It states that ``countries should consider the importance of a pro-innovation and proportionate governance and regulatory approach that maximises the benefits and takes into account the risks associated with AI. This could include making, where appropriate, classifications and categorisations of risk.''

The design challenge for regulation, to first approximation, is to determine which entities must comply with which obligations. 
How should policymakers structure regulation to tailor the obligations to the entities? 
Thresholds are a standard mechanism for determining the scope of regulation. 
Thresholds distinguish entities to allow for heightened obligations to be applied to some entities without being applied to others. We define a threshold as a simple decision rule where (i) a quantity of interest, such as model training compute measured in floating point operations (FLOP), is compared to a value, such as $10^{26}$ FLOP such that (ii) entities that surpass the threshold are subject to more regulatory scrutiny than those that do not. Thresholds often exempt certain entities from burdensome compliance to foster innovation. For example, businesses with less than \$500,000 in annual gross revenue do not meet the enterprise coverage threshold, exempting them from recordkeeping obligations under the U.S. Fair Labor Standards Act. Thresholds also narrow regulatory attention to a smaller set of entities that warrant greater focus or scrutiny. For example, U.S. banks with \$250 billion or more in total consolidated assets are designated as ``systemically important'' financial institutions, subjecting them to heightened scrutiny by the Federal Reserve under the Dodd-Frank Wall Street Reform and Consumer Protection Act.

While thresholds are ubiquitous across nearly every policy domain and jurisdiction, their design and application to AI policy remains nascent. As of January 2025, AI-specific thresholds have primarily been based on the computational cost of model training, which is measured in computational operations (\eg, FLOP). Noteworthy examples of compute thresholds include $10^{26}$ FLOP in the Biden Executive Order 14110 on Safe, Secure, and Trustworthy Development and Use of Artificial Intelligence and $10^{25}$ FLOP in the European Union’s AI Act. Other approaches have been considered: California’s proposed and vetoed SB 1047 Safe and Secure Innovation for Frontier Artificial Intelligence Models Act used the monetary cost of training, setting the threshold at \$100 million.

In general, thresholds are imperfect: Most risks vary continuously, whereas thresholds set discrete boundaries to distinguish among different entities. Thresholds are also imperfect proxies: By relying on specific, quantifiable metrics, they necessarily center what can be measured even where the underlying risk cannot be fully measured. Useful thresholds for frontier AI governance, therefore, must strike pragmatic compromises between (i) the theoretical constructs that should yield greater regulatory scrutiny and (ii) the practical operationalizations of what can be measured. 
Most fundamentally, the pace of technological change will require adaptive approaches that allow for thresholds to be flexible and readily updated to avoid ossification.\footnote{We focus on thresholds for foundation models but, in the design of AI, policymakers may want to differentiate the obligations for the initial in-scope developer vs. an entity downstream that modifies the model. Addressing derivative models, given the broad uptake of foundation models in the economy, along with more continual approaches to model development is essential for ensuring proportionality more broadly.}

\subsection{Thresholds for Foundation Models}
To proportionately regulate foundation models and the AI supply chain, policymakers must understand the many axes of variation to inform how they set thresholds. 
Supply chain monitoring \citep{bommasani2023ecosystem, cma2023ai} documents significant variation in the organizational profile of model developers, including their corporate status, employee head count, and geographic location. These developers expend very different amounts of resources to build their models, referring to costs in money, data, computation, and energy. 
The resulting models also differ greatly in fundamental properties like model architectures (\eg, dense models where all model parameters are active vs. sparse models where a small subset are active) and modalities (\eg, Meta’s unimodal text-to-text Llama 3.3 vs. Google’s highly multimodal Gemini 2.0 that can take text, code, images, audio, and/or video as inputs and generate text, audio, and/or images as output). They vary even more in their capabilities and associated risks when evaluated on standard benchmarks. Once development is completed, models vary greatly in how they are released (\eg, EleutherAI’s fully open Pythia, Meta’s open-weight Llama 3.3, OpenAI’s API-based GPT-4, Google’s fully closed Gopher). 
For all of these reasons and other business reasons, foundation models vary immensely in how they are used downstream and their ultimate impact on the public.

To define a threshold to distinguish foundation model developers, we see four natural approaches:
\begin{enumerate} 
\item \textbf{Developer-level properties:} 
For example, thresholding based on the number of employees could reduce obligations for small companies that may lack the personnel to comply. As an example from another policy domain, the Occupational Safety and Health Administration Recordkeeping and Reporting Occupational Injuries and Illnesses Standard partially exempts businesses with at most 10 employees from maintaining certain logs of work-related injuries and illnesses.
\item \textbf{Cost-level properties:} 
For example, thresholding based on the compute-related cost to develop a model could restrict regulatory attention to capital-intensive models given concerns of barriers to competition. As an example from another policy domain, the Clean Air Act’s program for the Prevention of Significant Deterioration requires permitting for facilities that emit, or have the potential to emit, either 100 tons per year of regulated air pollutant in specific source categories (\eg, chemical plants) or 250 tons per year of regulated air pollutant in all other source categories.
\item \textbf{Model-level properties:} 
For example, thresholding based on the model’s performance on benchmarks that involve identifying software vulnerabilities could identify models that may enable cyberattacks. As an example from another policy domain, the Consumer Product Safety Improvement Act uses laboratory testing for lead presence to limit lead to 100 parts per million of total lead content in any accessible part of children's products intended for children 12 years old and under.
\item \textbf{Impact-level properties:} 
For example, thresholding based on the number of commercial users of a model could identify models with broad dependence that may pose systemic risk. 
As an example from another policy domain, the European Union’s Digital Services Act imposes greater scrutiny on Very Large Online Platforms and Very Large Online Search Engines, defined as those with at least 45 million monthly active users in the EU.
Impact scales not only in the amount of usage, but also the type of usage (\eg the use of foundation models in high-impact AI systems that mediate life chances in domains such as employment, education, housing, health, and criminal justice).
\end{enumerate}

To determine the appropriate approach for setting a threshold in a given policy context, we recommend that policymakers adopt the default approach of using thresholds that align with the regulatory intent. For example, if the intent is to encourage up-and-coming startups, then shielding them via exemptions for small businesses may be appropriate: This intent could be expressed via a threshold based on business-level head count or revenue. If the intent is to discourage the misuse of certain dual-use capabilities, then categorizing entities based on evaluation results for their models for the associated capabilities may be appropriate. And if the intent is to place greater scrutiny on efforts likely generating large-scale disruption to societal operation and/or large-scale harm from the adoption of their technologies, then an impact-level designation may be appropriate.

While the principle of aligning the regulatory structure (\ie, the methods for thresholding) with the regulatory intent should be the starting point, practical concerns should also be considered. 
In practice, some of the quantities we describe above may be proxies for others.
In several cases, policymakers have already used training compute (a cost-level metric) as a proxy for some combination of greater capabilities, greater misuse risks, and/or greater downstream harm.

While a number of additional considerations influence threshold design \citep[see][]{bommasani2023tiers, heim2024trainingcomputethresholdsfeatures}, we highlight three:
\begin{enumerate}
\item \textbf{Determination time:} 
Different quantities first become measurable at different times. While a foundation model developer may estimate training compute even before training for internal budgeting purposes, the evaluation results for a model can only be determined after development and, still further, the downstream impact for a model can only be determined after deployment. In some cases, estimates can be prepared in advance, but these temporal effects have two key consequences. First, developers can better prepare to comply with policy if they are aware earlier that they will be in-scope. And second, certain obligations are difficult to impose if they require intervention before the determination time: It is difficult to expect a developer to implement certain governance practices for training data if they only determine they exceed a threshold after deployment.
\item \textbf{Measurability:}
Different quantities have differing complexity in how they are measured, spanning both the intrinsic difficulty of measuring the quantity and overarching standardization in how to measure the quantity. In particular, absent significant agreement on how to measure the quantity of interest, progress in implementation may stall as stakeholders disagree on the specifics of measurement.
\item \textbf{External verifiability:}
Different quantities vary in whether external parties of any kind are able to measure them. Prior AI policy in New York highlights that if the regulated entity is the one to unilaterally determine if they are in scope of the policy, then the policy may fail to have the desired effect due to widespread null compliance (Wright et al., 2024). For example, risk evaluation results may be externally reproduced to confirm developer’s designation, whereas training compute may require the ability to monitor the compute resources developers use, and training data size may be entirely unobservable given the lack of access to training data. 
\end{enumerate}

Since policy may have different regulatory intents and existing thresholds vary in their profiles of determination time, measurability, and external verifiability, we agree with \citet{nelson2024ntia} that ``a one-size-fits-all approach or a single threshold metric is inadequate for governance because different AI systems and their outputs present unique challenges and risks.'' To this end, we point to the European Union’s AI Act, which designates models trained with $10^{25}$ FLOP as posing systemic risk as of March 2025 as the default criteria. 
However, the AI Act in Annex XIII affords the regulator flexibility to also consider alternatives metrics, such as the number of parameters, size of the data set, estimated cost or time of training, estimated energy consumption, benchmarks and evaluations of capabilities of the model, and whether the model has a high impact on the internal market due to its reach (either due to at least 10,000 registered business users or the number of registered end users). Further, to capture fast-moving scientific developments, the AI Act creates a scientific panel that is empowered to issue qualified alerts to identify models that may pose systemic risk even if they are not captured by predefined quantitative thresholds.

Overall, we emphasize that irrespective of the combination of metrics deemed most appropriate in the present, policymakers should ensure that mechanisms exist not only to update specific quantitative values, given the rapid pace of technological and societal change in AI, but also to change the metrics altogether.

\subsection{Guiding Principles}
\textit{Generic developer-level thresholds seem to be generally undesirable given the current AI landscape.} Since many small entities can develop hugely influential and potentially risky foundation models, as demonstrated by the Chinese company DeepSeek, the use of thresholds based on developer-level properties may inadvertently ignore key players. 
In fact, major players in the space, such as Anthropic, OpenAI, and xAI, may be relatively small according to some conventional metrics of businesses (\eg, head count). 
At the same time, these approaches may bring into scope massive, established companies in other industries that are simply exploring the use of AI since thresholds based on properties of companies may not distinguish between the entire business and the AI-specific subset. 
Therefore, we caution against the use of customary developer-level metrics that do not consider the specifics of the AI industry and its associated technology. \\

\noindent \textit{Cost-level thresholds such as training compute should not be used alone.}
Training compute is by far the most common metric for thresholds in existing AI policy given its clear strengths \citep{heim2024trainingcomputethresholdsfeatures}: 
(i) Fairly authoritative methods exist for measuring compute \citep{fmf2024flops, casper2025practicalprinciplesaicost}; (ii) Compute can be estimated prior to the development of the model; and (iii) Compute may be externally verifiable by entities that monitor or operate compute facilities.
However, training compute has clear deficits that are well-established:
(i) Models across different modalities currently use vastly different amounts of training compute;
(ii) Thresholds for training compute may need to be updated very frequently due to rapid improvements in algorithmic efficiency;
and (iii) Training compute has limited predictive power as a proxy for risk because it does not factor in how models are distributed and adopted in the economy \citep{nelson2024ntia, bommasani2023tiers, hooker2024limitationscomputethresholdsgovernance}.
These concerns are exacerbated by recent developments such as (iv) the growing shift from training-time compute allocation to inference-time compute allocation and (v) the complex accounting required to address rapidly evolving technical practices such as model distillation and the use of large-scale synthetic data.

On this basis, we conclude that if training compute thresholds are used at all, they may function best when used as an initial filter to cheaply screen for entities that may warrant greater scrutiny \citep{nelson2024ntia, bommasani2023tiers, heim2024trainingcomputethresholdsfeatures, hooker2024limitationscomputethresholdsgovernance}. 
More broadly, within the family of cost-level metrics, training compute may still be the most attractive option, given other options are difficult to calibrate across different developers (\eg dataset size, monetary cost), but they should actively monitor the rapid changes in how compute is allocated toward foundation model development and deployment.\\

\noindent \textit{Thresholds based on risk evaluation results and observed downstream impact are promising for safety and corporate governance policy, but they have practical issues.} Policy that aims to manage risk by influencing developer practices often addresses one or both of (i) misuse of foundation models by malicious actors and (ii) sociotechnical harms from widespread deployment. Therefore, these intents align directly with evaluations of risk (\eg, model performance at generating child sexual abuse material or cyberattacks) as well as the downstream footprint of models (\eg, if models are integrated into high-stakes decision-making systems for hiring or benefits determination). 

However, the question of which risk evaluations are sufficiently trustworthy looms in the air: The pre-deployment testing conducted by the U.S. AI Safety Institute and U.K. AI Security Institute may provide some guidance on this. In addition, industry-led safety frameworks, which articulate explicit thresholds of risk based on evidence of measured capabilities, may offer a pathway for industry consensus. Agreement on which risks should be tracked and how they can be measured through evaluations, ideally harmonized to level the playing field through third-party auditing, offers an industry-inspired pathway that can inform governance.

On downstream impact, model developers may not have this information, especially in the case of open models, so other mechanisms for market monitoring may be required instead. Overall, policy can actively develop the infrastructure to turn these approaches into fully viable options in threshold design.

\hypertarget{Feedback}{\section{Summary of Feedback and Changes}}
\label{sec:feedback}
A draft of this report was shared publicly online in March 2025, along with a form (\autoref{sec:feedback-form}) to encourage feedback and input from a wide range of experts. 
The Working Group deeply appreciates the individuals and organizations from numerous disciplines and sectors who participated by sharing valuable insights, references, and suggestions, including, in some cases, by making their responses public to contribute to the broader discourse. 
We received over 60 feedback submissions, which supplement the Working Group’s internal review process and initial round of external expert review (see Acknowledgements on pg. \pageref{sec:acknowledgments}). 
Three sectors (academia, industry/business/private sector, and NGO/civil society) contributed the vast majority of feedback, with each contributing at least 25\% of all submissions.

The feedback reflects a range of views and fundamental questions that can inform California AI policy and governance. Feedback addressed clarifications on report content, critiques of substantive content, and future directions for further inquiry. 
We summarize this feedback below without individual-level attribution, consistent with the privacy expectations in the feedback form (\autoref{sec:feedback-form}).
Finally, we explain how we incorporated this feedback into the final report.

\subsection{Clarifications on Report Content} 
Several feedback responses indicated the desire for clarity on concepts and definitions introduced in the draft report. Other feedback indicated that the overarching remit of the report was unclear.\\

\noindent \textit{Clarity on the concept of evidence-based policy.} 
Comments indicated confusion on the overarching frame of evidence-based policy and the discussion of evidence-generating policy. Often this confusion rested on either (i) how evidence-based policy is understood given current incentives that yield the present information asymmetries between foundation model developers, governments, and the public or (ii) how evidence contributes to improved accountability.\\

\noindent \textit{Clarity on the current state of AI.} 
Comments identified current developments in the AI ecosystem that went beyond the report’s description of current capabilities, risks, and practices. In particular, several of these comments addressed the current evidence for risks, the growth of safety frameworks and cases, and the adherence of companies to voluntary commitments.\\

\noindent \textit{Clarity on the specific definitions.} 
Comments on definitions primarily indicated confusion around the use of several closely related terms to describe current AI technologies (\eg, frontier AI, foundation models, artificial general intelligence), though other feedback appreciated the granularity of the terminology. Other comments addressed the more general definitions for artificial intelligence, AI model, and AI system, as well additional jargon (\eg, marginal risk, see \autoref{sec:intro-risks}).\\

\noindent \textit{Clarity on the remit for the report.} 
Comments often expressed uncertainty about or provided recommendations for the report scope, the report purpose, and the Working Group remit. More specifically, several comments addressed the unique position of impact California holds in the ecosystem as the home to many leading AI companies, a significant fraction of the AI workforce, and past governance initiatives on digital technology. Several comments, which were notably received prior to recent developments at the federal level, discussed the importance of California to the U.S. AI ecosystem, advocating to avoid overregulation and instead create positive incentives for companies to compete on safety. Other comments addressed concerns of fragmentation with efforts in other jurisdictions, such as existing and ongoing legislative efforts in other states, and opportunities for harmonization, including with the EU AI Act and international standards. Finally, some comments touched on the impact of California’s AI policy on U.S. national competitiveness and the broader geopolitical environment.

\subsection{Critique of Substantive Content}
The draft report’s core sections received significant support, which we do not summarize for brevity. Instead, we summarize targeted critiques, which we highlight as valuable alternative perspectives. Although the Working Group maintains different views on many of these critiques, they offer important alternative perspectives that enhance the overall discussion and merit continued public discourse.\\

\noindent \textit{Critique of context (\autoref{sec:context}).} 
Some comments indicated that AI companies take more voluntary action than what is seen in the discussed cases: They promote AI safety and transparency by publishing safety reports and systems cards, releasing open-weight models, issuing responsible scaling frameworks, soliciting and managing red-teaming efforts, and developing model documentation standards. Therefore, these companies are more cooperative and proactive in safety compliance compared to tobacco and fossil fuel companies. Other comments suggested that the cases generally describe policy successes, but not policy failures, and proposed other cases that could inform California’s AI policy.\\

\noindent \textit{Critique of transparency (\autoref{sec:transparency}).} 
Several comments indicated that transparency should be purposeful by clarifying how disclosures relate to accountability and being aware of the incentives that shape the quality of the disclosed information. Some comments recommended focusing on specific implementation rather than broader guidance, on more restricted non-public disclosures including via redaction, and on the costs of transparency. Specific attention focused on developing the discussion of whistleblower protections, including their relationship to existing protections. Specific attention was also focused on third-party evaluations (sometimes referred to as audits in the feedback) by highlighting the immature third-party evaluation ecosystem, the lack of evaluation standards and broader measurement science, and the current barriers for evaluators to acquire access to the AI models and systems they evaluate.\\

\noindent \textit{Critique of adverse event reporting (\autoref{sec:aer}).} 
Several comments questioned the absence of precise definitions for key terms (\eg, adverse event, incident, harm, unacceptable behavior) or the detailed process for adverse event reporting. Some comments identified practical challenges as both hurdles for a statewide reporting database and costs for reporting entities. Others described more participatory approaches to reporting to maximize data collection while minimizing compliance burden, including via collaboration with existing efforts.\\

\noindent \textit{Critique of scoping (\autoref{sec:scoping}).} 
Many comments emphasized the importance of adaptive thresholds, indicating consensus that thresholds cannot be static but change to reflect evolving context. Many comments specifically addressed compute thresholds and expressed diverging views: Some 
argued compute thresholds are useful and necessary, while others argued for a stronger dismissal of compute thresholds given their limits as a proxy for risk, though all comments acknowledged that compute thresholds have significant flaws. Some comments addressed thresholds we did not discuss, such as those based on more qualitative criteria, or the broader relationship between scoping thresholds and policy interventions in the form of red lines.

\subsection{Future Directions for Further Inquiry}
The draft report’s scope was determined by the Working Group’s interpretation of the request from California Governor Gavin Newsom. 
We summarize feedback on out-of-scope topics.\\

\noindent \textit{Downstream AI applications.}
Many comments discussed specific application areas for foundation models as part of the increasing deployment of AI across economic sectors. Some comments advocated for use case-based regulatory efforts, especially in domains that pose critical risks or use methods customary to product safety, while other comments advocated for greater focus on specific deployment contexts and resulting AI systems. The feedback also discussed the need to better understand the economic impacts of AI within specific downstream domains, including to understand the implications for the future of work and the labor economy.\\

\noindent \textit{The AI talent pipeline.}
Several comments discussed the need to focus on preparedness for AI developments across the state’s population, especially for workers and students. These comments suggested initiatives to improve workforce training for workers across the AI supply chain, including via collaboration with the state’s educational institutions. Separately, other comments discussed the spillover effects of AI on educational pipelines, questioning the impacts on the quality of student education in the presence of widespread AI technology and discussing the procurement opportunities for AI technologies within the education sector.\\

\noindent \textit{Resourcing the AI ecosystem.}
Several comments stressed the importance of public funds to advance research on AI, highlighting the strengths of California’s research institutions, the initiatives in other states such as New York’s Empire AI, and the deficits in access to compute for researchers. A number of comments more broadly discussed the resources for the entire innovation ecosystem, including some that specifically focused on venture capital and the importance of policy as a fuel for entrepreneurial startups. Other comments discussed how initiatives to resource the ecosystem could be used to implement more responsible practices.\\

\noindent \textit{The governance of open and closed models.}
Many comments mentioned the distinction between open and closed models in the AI ecosystem. Some highlighted that the open release of foundation models, including via broader open-source development practices, is critical for both spurring innovations and increasing safety and transparency. Others highlighted the risks of open models. Overall, many comments on this topic recognized both the benefits and risks of openness, referencing the National Telecommunications and Information Administration report \citep[NTIA;][]{ntia2024open} as a key starting point for governance on this topic. \\

\noindent \textit{Regulatory capacity on AI.} 
Several comments raised concerns about California’s capacity to regulate AI, highlighting challenges with past digital technology policy. These comments questioned the technical expertise on AI within the government and described pathways for improving state capacity. Other comments indicated that current regulatory capacity may render initiatives to design guardrails ineffective or outright counterproductive: Many of these comments focused on flexible policy design, but a few expressed pessimism about the viability of meaningful oversight.

\subsection{Changes to the Report}
Based on further review, feedback, and the rapid pace of change in the field of frontier AI, the authors revised the draft report to produce the final report.
We summarize the substantive changes.\\

\noindent \textit{Changes to introduction (\autoref{sec:introduction}).}
This section was updated to provide a more rigorous discussion of the relationship between state policy on frontier AI and policy in other jurisdictions, especially the federal government. In addition, the updated section articulates how California frontier AI policy can navigate challenging geopolitical realities, leveraging excellent quality and safety standards to strengthen national competitiveness. Finally, multiple paragraphs which discuss new empirical evidence for frontier AI risks were added.\\

\noindent \textit{Changes to context (\autoref{sec:context}).}
This section was updated to include a more rigorous discussion of regulatory incentive structures from other cases, including both positive inducements and deterrents. The updated section more directly addresses the relationship between state and federal regulation in historical context, including updating multiple case studies with relevant state-federal dynamics. The revised section emphasizes building safety into initial technological frameworks through proactive risk assessment and mitigation strategies. Further changes include greater emphasis on aligning incentives for transparency and accountability not just for frontier AI developers but all relevant actors in the ecosystem, building evidence-generating policy that accounts for internal AI deployments, and articulating how policy can help industry navigate competing incentives, all of which enhance the foundation for a “trust but verify” approach. \\

\noindent \textit{Changes to transparency (\autoref{sec:transparency}).}
This section was updated to strengthen the discussion of openness in relation to transparency, acknowledge First Amendment challenges to mandatory disclosures, and emphasize that transparency approaches are most valuable when they are purposeful, actionable, and outcome focused. Additional updates include references to regulatory harmonization, post-deployment economic monitoring, safety cases, whistleblower protections for non-employees, how states could leverage the existing technical infrastructure and expertise, and clarification that safe harbors discussed relate to third-party researchers and not AI companies.\\

\noindent \textit{Changes to adverse event reporting (\autoref{sec:aer}).}
This section was updated to further emphasize the benefits of an initially narrow definition for adverse events to avoid a deluge of reports that masks the signal in the reporting system. 
Additional updates include mentioning the possibility for reporting fatigue and immunity provisions in the Aviation Safety Reporting System.\\

\noindent \textit{Changes to scoping (\autoref{sec:scoping}).}
This section was updated to strengthen the critique of thresholding based exclusively on training compute and to emphasize the need for updating given rapid technological change. Additional updates include mentions of impact beyond reach, distillation, algorithmic efficiency, and the complexities of proportionately addressing model derivatives.

% In addition, clarificatory edits have additionally been made throughout the report. 
\clearpage
\bibliographystyle{acl_natbib}
\bibliography{refdb/all, main}
\clearpage
\appendix
\section{Feedback Form}
\label{sec:feedback-form}
Following the release of the draft report, we created a feedback form to receive public input. 
We provide a copy of this form below.

\subsection{Preamble Text}
In September 2024, Governor Gavin Newsom announced \href{https://www.gov.ca.gov/2024/09/29/governor-newsom-announces-new-initiatives-to-advance-safe-and-responsible-ai-protect-californians/}{new initiatives} to advance safe and responsible artificial intelligence (AI)
 development in the State. As part of that announcement, Governor Newsom requested that Dr. Fei-Fei Li, Co-Director of the Stanford Institute for Human-Centered Artificial Intelligence (HAI); Dr. Mariano-Florentino Cuéllar, President of the Carnegie Endowment for International Peace; and Dr. Jennifer Tour Chayes, Dean of the UC Berkeley College of Computing, Data Science, and Society, prepare a report focusing on frontier models. Note: the Draft Report represents scholarly work from each of the Co-Leads and does not represent the positions of their institutions.

To encourage feedback and input from a wide range of expertise, the Draft Report is being shared publicly. We welcome feedback, reflections, and additional information from a variety of disciplines and sectors by \textbf{April 8, 2025} for consideration in advance of the final report. 

The working group will not share individual responses to this form publicly, nor attribute any feedback or input at an individual level. The feedback will be aggregated and incorporated into the final report, expected by June 2025. Thank you!

\subsection{Information about Respondent}
\begin{itemize}
\item Email Address \textit{(Required)}
\item First Name \textit{(Optional)}
\item Last Name \textit{(Optional)}
\item Please check the box if you would like to opt in for receiving email updates. [Checkbox] \textit{(Optional)}
\item Which category best describes you? [Select one] \textit{(Optional)}
\subitem Academia
\subitem Government/Public Sector
\subitem NGO/Civil Society
\subitem Industry/Business/Private Sector
\subitem Other [free response field provided]
\item Title \textit{(Optional)}
\item School/Company/Organization/Agency Name \textit{(Optional)}
\end{itemize}

\subsection{Form Questions}
\begin{enumerate}
\item The structure of the Draft Report includes sections about \textit{Context, Transparency and Third-Party Risk Assessment, Adverse Event Reporting, and Scoping}. At a high level, what might you find valuable? What types of questions are you most interested in, and how might you use the report in your work? \textit{(Optional)}

\item From your perspective and experience, what key factors do you see affecting California’s path forward in AI governance? Please feel free to provide specific feedback referring to the sections of the draft report. \textit{(Optional)}

\item Numerous frontier AI governance-focused groups have been working on frameworks, guidance, and reports aiming to leverage scientific research. For what topics or issues are you observing challenges in reaching scientific consensus? Do you have recommendations to bridge gaps? \textit{(Optional)}

\item What could be done individually or collectively to leverage frontier AI for Californians' benefit? \textit{(Optional)}

\item Please feel free to list any published resources you would like to share with the Joint California Policy Working Group on AI Frontier Models. \textit{(Optional)}
\end{enumerate}

\end{document}